\documentclass[aps,a4paper,superscriptaddress,nofootinbib,twocolumn]{revtex4-2}
\usepackage{amsmath,amssymb,graphicx,multirow,dcolumn,bm,latexsym,soul,nicefrac}
\usepackage[colorlinks,linkcolor=blue,citecolor=blue,urlcolor=blue ]{hyperref}
\usepackage{acronym}
\usepackage{subfigure}
\usepackage{appendix}
\usepackage{longtable}
\usepackage{footnote}
\usepackage{tabularx}
\usepackage{siunitx}
\usepackage{float}

\newcommand{\SYSU}{School of Science, Shenzhen Campus of Sun Yat-sen University, Shenzhen 518107, China}
\newcommand{\ASTU}{School of Electrical and Electronic Engineering, Anhui Science and Technology University, Bengbu, Anhui 233030, China} 

\begin{document}
\title{Testing Gravitational-Wave Signal From Verification Binaries with Space-Based Gravitational-Wave Detectors}

\author{Zi-Heng Yu}
\email{These authors contributed equally to this work and should be considered co-first authors.}
\affiliation{\SYSU}

\author{Sen Yang}
\email{These authors contributed equally to this work and should be considered co-first authors.}
\affiliation{\SYSU}

\author{Liangliang Ren}
\affiliation{\ASTU}

\author{Shun-Jia Huang}
\email{corresponding author: huangshj69@sysu.edu.cn}
\affiliation{\SYSU}

\date{\today}

\begin{abstract}
Space-based \ac{GW} detectors will open the millihertz band to survey \acp{UCB}. 
\textit{Verification binaries} (VBs) is a key to verifying the performance of space-based \ac{GW} detectors because its parameters are known from electromagnetic observations and it is expected to be a detectable source of \ac{GW}.
We evaluated 72 VBs, computing their detection prospects and parameter estimation precision for individual \ac{GW} detectors and networks. 
Among single detectors, DECIGO shows the highest sensitivity, detecting 71 sources at signal-to-noise ratio $\rho$ $\geq$ 5, compared to 42 for LISA, 32 for Taiji, and 27 for TianQin, while the full TianQin + LISA + Taiji + DECIGO network improves this to 72 detectable sources.
For parameter estimation, individual detectors achieve median precisions on the order of $\sim 10^{-2}-10^{-1} \, \text{M}_{\odot}$ for chirp mass, $\sim 1\,\text{kpc}$ for distance, $\sim 1-3.4\,\text{deg}$ for inclination and $\sim 10^{-4}-10^{-1}\,\text{deg}^2$ for sky localization.
The complete TianQin + LISA + Taiji + DECIGO network enhances these constraints substantially, reducing the median uncertainties to approximately $\sim 10^{-2} \, \text{M}_{\odot}$ in chirp mass, $\sim 10^{-1}\,\text{kpc}$ in distance, $\sim 1.6\,\text{deg}$ in inclination and $\sim 10^{-4}\,\text{deg}^2$ in sky localization.
The upcoming space-based \ac{GW} detectors, especially their networks, have outstanding observational capabilities for \ac{UCB}, which will advance our research on multi-messenger astronomy and deepen our understanding of \ac{UCB} in the Milky Way.

\end{abstract}

\maketitle
\acrodef{GW}{gravitational wave}
\acrodef{EM}{electromagnetic}
\acrodef{aLIGO}{Advanced LIGO}
\acrodef{aVirgo}{Advanced Virgo}
\acrodef{UCB}{ultra-compact binarie}
\acrodef{DWD}{double white dwarf binarie}
\acrodef{AM CVn}{AM Canum Venaticorum binarie}
\acrodef{CWDB}{close white dwarf binarie}
\acrodef{UCXB}{ultra-compact X-ray binarie}
\acrodef{sdB}{Subdwarf binarie}
\acrodef{WD}{white dwarf}
\acrodef{NS}{neutron star}
\acrodef{BH}{black hole}
\acrodef{MS}{main-sequence}
\acrodef{SNR}{signal-to-noise ratio}
\acrodef{FIM}{Fisher information matrix}
\acrodef{BBH}{binary black hole}
\acrodef{BNS}{binary neutron star}
\acrodef{AGN}{active galactic nuclei}
\acrodef{EMRI}{extreme mass ratio inspiral}
\acrodef{SNe Ia}{type-Ia supernovae}
\acrodef{PSD}{power spectral density}
\acrodef{CVB}{candidate \textit{verification binarie}}
\acrodef{VB}{\textit{verification binarie}}
\acrodef{GR}{general relativity}

\section{Introduction}\label{sec:1}

In 1916, based on \ac{GR}, Albert Einstein predicted the existence of \acp{GW} \cite{1916SPAW.......688E}.
\acp{GW} propagate the perturbations of spacetime curvature, radiating outward at the speed of light. 
These waves convey energy and angular momentum, and critically encode detailed astrophysical information regarding the dynamics of their sources \cite{RevModPhys.52.299,1982ApJ...253..908T}. 
As such, they constitute a qualitatively new messenger, distinct from and complementary to \ac{EM} radiation \cite{1986Natur.323..310S}.
Currently, the prevailing detection modality is laser interferometry \cite{2015CQGra..32g4001L,2015CQGra..32b4001A,2020LRR....23....3A}, specifically the Michelson interferometric architecture. 

In 2015, \ac{aLIGO} \cite{2015CQGra..32g4001L} made the first direct detection of \ac{GW} (GW150914) \cite{2016PhRvL.116f1102A}, initiating a new era in \ac{GW} astronomy.
In the following years, \ac{aVirgo} \cite{2015CQGra..32b4001A} and KAGRA \cite{KAGRA:2012} also operated successively to detect \ac{GW} signals, forming the LIGO-Virgo-KAGRA (LVK) detection network together with \ac{aLIGO}.
So far, LVK has detected $\sim$ 90 \ac{GW} events \cite{Abbott:2019,Abbott:2021a,Abbott:2021b,Abbott:2023,Abbott:2024,Abac:2024}, which made the GW observation a new window to understanding the Universe.

Ground-based interferometers such as \ac{aLIGO} and \ac{aVirgo} are sensitive only to the $10^1-10^3$ Hz band \cite{2019arXiv190304592M}, but \ac{GW} span many orders of magnitude in frequency, requiring complementary facilities: the cosmic microwave background polarization experiments for femtohertz \cite{annurev:/content/journals/10.1146/annurev-astro-081915-023433}, pulsar-timing arrays for nanohertz \cite{2018ApJ...859...47A,2015Sci...349.1522S}, and space-based laser interferometers for millihertz \cite{2017arXiv170200786A,2016CQGra..33c5010L}.
The millihertz band hosts the most diverse population: massive black-hole binaries ($10^3-10^7 \,\text{M}_\odot$) from galaxy mergers \cite{PhysRevD.93.024003,1998AJ....115.2285M,1969Natur.223..690L}, extreme-mass-ratio inspirals (EMRIs) \cite{2017PhRvD..95j3012B,PhysRevD.102.063016}, and \acp{UCB} of \acp{WD}, neutron stars, and stellar-mass black holes \cite{2018MNRAS.480.2704L,2017MNRAS.470.1894K,2018PhRvD..98f4012R,2020MNRAS.492.3061L}, together with astrophysical and cosmological stochastic backgrounds \cite{2017LRR....20....2R}, thus bridging astronomy and fundamental physics \cite{2019NatAs...3..858T,PhysRevD.71.084025}.
Among these \ac{GW} sources, \acp{UCB} have the highest detectable quantity, and space-based laser interferometers are expected to capture tens of thousands of \acp{UCB} \cite{2001A&A...375..890N,2010A&A...521A..85Y,2017MNRAS.470.1894K,2020PhRvD.102f3021H}.

If \ac{GW} detectors can also detect the \ac{GW} signals radiated by \acp{UCB} that \ac{EM} telescopes have observed, scientists can use such sources, the so-called \acp{VB}, to calibrate the instrument noise of \ac{GW} detectors themselves.
In 2006, A. Stroeer and A. Vecchio et al. \cite{2006CQGra..23S.809S} performed a comprehensive parameter analysis of 30 verification-binary candidates, focusing on the \acp{SNR} expected for these systems. 
They concluded that RX J0806.3+1527, V407 Vul, ES Cet, and AM CVn satisfy the criteria for classification as reliable \acp{VB}.
In 2018, Kupfer et al. \cite{2018MNRAS.480..302K} computed the \acp{SNR} of approximately fifty \acp{VB} using parallaxes from Gaia Data Release 2. 
For the systems exhibiting high \acp{SNR}, they employed \ac{FIM} formalism to forecast the parameters uncertainty on the \ac{GW} amplitude ($\mathcal{A}$) and the inclination angle ($\iota$).
In 2020, Shun-Jia Huang et al. \cite{2020PhRvD.102f3021H} investigated the detection prospects for Galactic \acp{DWD} with the space-borne \ac{GW} observatory TianQin. 
They concluded that TianQin is expected to resolve 12 of the $\sim100$ \acp{DWD} already identified in the Galaxy, and TianQin could expand the \ac{DWD} catalogue to approximately $10^{4}$. 
The authors further emphasized the scientific benefits of coordinated multi-messenger campaigns combining TianQin data with \ac{EM} facilities.

Recently, scientists have discovered and reported more galactic binaries.
Kilic et al. \cite{2021ApJ...918L..14K} reported the discovery of SMSS J0338-8139 and SDSS J0634+3803, which could be verification binaries for LISA.
Chandra et al. \cite{2021ApJ...921..160C} reported SDSS J1337+3952, which will be detectable by future space-based interferometers.
Ren et al. \cite{2023ApJS..264...39R} represented a binary sample of 429 \acp{CWDB} candidates and calculate the corresponding \ac{SNR} for TianQin and LISA.
Burdge et al. \cite{2023ApJ...953L...1B} reported ZTF J0127+5258, which contains a white dwarf arrector and a warm donor.
They predict the source will be detected by LISA after $4$ years of observations.
Kupfer and Korol et al. \cite{2024ApJ...963..100K} updated 55 sources and calculated distances.
Using Bayesian analysis methods, they predicted the LISA detectability for these sources.
Kosakowski et al. \cite{2025ApJ...987..205K} reported SMSS J1138-5139 and expected LISA will detect its \acp{GW}.
In our research, we add \acp{CWDB}, \acp{sdB} and \acp{UCXB} to our catalogue.

Building on previous research, this paper updates the \ac{VB} catalogue and analyzes the detection capabilities of the individual space-based \ac{GW} detectors and joint detection by multiple detectors.
In Section \ref{sec:2}, we outline the sample of the currently known \ac{VB}.
In Section \ref{sec:3}, we derive formulas for computing the \ac{SNR} and uncertainties on binary parameters.
In Section \ref{sec:4}, we calculate the detection numbers and parameter estimation of nine configurations  involving four individual detectors and five detector networks for \acp{VB}.
Finally, we summarize our main findings in Section \ref{sec:5}.

\section{Verification  Binaries}\label{sec:2}

We assembled a sample of five classes of \ac{VB} systems: \acp{DWD}, \acp{AM CVn}, \acp{CWDB}, \acp{UCXB} and \acp{sdB}.

\acp{DWD} constitute the most abundant class of \acp{UCB} in the Milky Way.
They are expected to be the principal contributor to the signals detected by space-based \ac{GW} detector \cite{1990ApJ...360...75H}.
Millions of \acp{DWD} radiate low-frequency \ac{GW} signals, merging into an irresolvable background noise or the galactic background. 
In contrast, the louder, individually resolvable sources emerge above this confusion limit; together with high-frequency sources, these objects constitute a population of $\sim10^{4}$ detectable white-dwarf binaries \cite{2001A&A...375..890N,2010A&A...521A..85Y}.
\acp{DWD} can be assembled from helium-core, carbon/oxygen-core, or oxygen/neon/magnesium-core \acp{WD} in any combination, with a combined mass typically below $1.4\, \text{M}_\odot$ \cite{2023LRR....26....2A}.

\acp{AM CVn}, which originate from \acp{DWD}, consist of a \ac{WD} accreting matter from a hydrogen-deficient star or \ac{WD} companion \cite{1995Ap&SS.225..249W,2010PASP..122.1133S}.
\acp{AM CVn} are of great value to space-based \ac{GW} detectors, as they deliver multi-band \ac{EM} information and their \ac{GW} frequencies fall mainly within the sensitive
bands of space-based \ac{GW} detectors \cite{2025RPPh...88e6901L}.
Typical \ac{AM CVn} systems such as J0806, V407 Vul and ES Cet have been highlighted as important verification sources for space-based \ac{GW} detectors (e.g. \cite{2018MNRAS.480..302K,2020PhRvD.102f3021H}).

\acp{CWDB} are compact binary systems in which at least one component is a \ac{WD}.
They are an important branch of the evolution channel of \ac{MS} star binaries.
Depending on the component type and the mass-transfer status, \acp{CWDB} are usually divided into three main classes: post-common-envelope binaries, detached systems composed of a \ac{WD} and a \ac{MS} star \cite{2023ApJS..264...39R,2016PASP..128h2001H}; cataclysmic variables, semi-detached systems in which a low-mass donor fills its Roche lobe and transfers matter to the \ac{WD} \cite{2011ApJS..194...28K}; and \acp{DWD}, detached pairs of two \acp{WD} that are the end products of two successive common-envelope phases \cite{2001A&A...375..890N}. 
The orbital periods of \acp{CWDB} range from a few minutes to a day, with the shortest-period objects ($P_{\mathrm{orb}} \lesssim 60$ min) being the most potential individually resolvable \ac{GW} sources for millihertz \ac{GW} detectors such as LISA and TianQin \cite{2023ApJS..264...39R}.
ZTF J052610.40+593445.85 is a \ac{CWDB} with shortest orbital period of $20.5$ minutes and constitutes a potentially detectable millihertz \ac{GW} source \cite{2023ApJS..264...39R}.
  
\acp{UCXB} comprise a neutron star or stellar-mass black hole plus a donor star, with typical orbital periods of less than $60$ min \cite{1982ApJ...254..616R,2023A&A...677A.186A}.
The compactness of \ac{UCXB} confines the donor to be \ac{WD}, semi-degenerate dwarf, or helium star \cite{1982ApJ...254..616R}.
\acp{UCXB} are also important \ac{GW} sources.
1RXS J171824.2-402934 ($P_{\mathrm{orb}} \sim 7.0$ min), 4U 1820-30 ($P_{\mathrm{orb}} \sim 11.4$ min) and 4U 0513-40 ($P_{\mathrm{orb}} \sim 16.9$ min) are all candidate sources within the millihertz band of space-based detectors \cite{2009A&A...506..857I,2020ApJ...900L...8C,2024ApJ...963..100K}.

\acp{sdB} contain a subdwarf B star together with a compact companion, a \ac{WD} or neutron star.
Angular-momentum loss via \ac{GW} combined with magnetic braking shrinks the orbit further, leading a \ac{sdB} to a \ac{UCXB} or a millisecond-pulsar binary \cite{2023LRR....26....2A}.
The most compact \acp{sdB} attain orbital periods $P_{\mathrm{orb}} \lesssim 1$ hour \cite{2012ApJ...759L..25V, 2013A&A...554A..54G, 2017ApJ...851...28K, 2020ApJ...898L..25K}, which makes them potential detectable sources.
OW J0741 ($P_{\mathrm{orb}} \sim 45.0 $ min), CD-30 11223 ($P_{\mathrm{orb}} \sim 70.9 $ min) and HD 265435 ($P_{\mathrm{orb}} \sim 99.1 $ min) are \acp{sdB} whose orbital frequencies fall in the millihertz band.
They are potentially detected by the space-based \ac{GW} detectors \cite{2017ApJ...851...28K,2013A&A...554A..54G,2021PTEP.2021eA107M}.

We have collected the currently known candidates of these five \ac{VB} types and recorded their measured and estimated \ac{GW} parameters in Table \ref{tab:sample} (Appendix \ref{app:parameters}).
This source table includes $20$ \acp{AM CVn}, $33$ \acp{DWD}, $13$ \acp{CWDB}, $3$ \acp{UCXB} and $3$ \acp{sdB}.

\section{modeling signal and noise}\label{sec:3}
\subsection{Gravitational wave signal from a single binary} \label{sec:3a}

For each binary, it can be described by the following parameters: orbital period $P$, component masses $m_1$ and $m_2$, the ecliptic longitude $\lambda$ and latitude $\beta$, distance from the Sun $d$, inclination angle $\iota$, polarization angle $\psi_S$, and the initial orbital phase $\phi_0$.

For binaries whose frequencies are significantly smaller than the mission lifetime of many detectors of several years, they can be safely considered as monochromatic sources, meaning that the \acp{GW} generated by them can be described by a set of seven parameters: the dimensionless amplitude $\mathcal{A}$, \ac{GW} frequency $f=2/P$, $\lambda$, $\beta$, $\iota$, $\psi_S$, and $\phi_0$.

From the quadrupole formula of \ac{GW}, the waveforms of the plus + and cross × modes in the principal polarization coordinate are given  \cite{PhysRev.131.435,Landau:1962}
\begin{equation}
h_+(t) = \mathcal{A}(1+\cos\iota^2)\cos(2\pi ft + \phi_0 + \Phi_D(t))\,,
\label{eq:hplus}
\end{equation}
\begin{equation}
h_{\times}(t) = 2\mathcal{A}\cos\iota\sin(2\pi ft + \phi_0 + \Phi_D(t))\,,
\label{eq:hcross}
\end{equation}
with
\begin{equation}
\mathcal{A} = \frac{2(G\mathcal{M})^{5/3}}{c^4d}(\pi f)^{2/3},
\label{eq:GWamp}
\end{equation} 
where $\mathcal{M}\equiv (m_1 m_2)^{3/5}/(m_1+m_2)^{1/5}$ is the chirp mass, and $G$ and $c$ are the gravitational constant and the speed of light, respectively.
The additional term $\Phi_D(t)$ in the \ac{GW} phase [Eq.~\eqref{eq:hplus}-\eqref{eq:hcross}] is the Doppler phase arising from the periodic motion of detector around the Sun:
\begin{equation} 
\Phi_D(t) = 2\pi f t \frac{R}{c} \sin(\pi/2-\beta) \cos(2\pi f_m t-\lambda)\, ,
\label{}
\end{equation}
where $R=1\mathrm{A.U.}$ is the distance between the Earth and the Sun, and $f_m=1/\mathrm{year}$ is the modulation frequency. 

\subsection{Detector response} 
\label{sec:3b}

The prevailing space-based \ac{GW} detectors envision a unit of three drag-free satellites orbiting the Earth or the Sun.
Satellites will form an equilateral triangle unit oriented in such a way that the normal vector to the detector’s plane is pointing towards a specific source or cyclic.

The triangle unit can be regarded as the two Michelson interferometers 1 and 2. 
For each interferometer, it can record the \ac{GW} strain as the detector \cite{PhysRevD.57.7089}:
\begin{equation} h(t)=h_+(t)F^+(t,\theta_S,\phi_S,\psi_S)+h_{\times}(t)F^\times(t,\theta_S,\phi_S,\psi_S)\,,
\label{eq:waveform_1}
\end{equation}
here $F_{+,\times}$ are the antenna pattern functions. 
($\theta_S$,$\phi_S$) is the latitude and longitude of the source in the detector’s coordinate frame.
The antenna pattern functions can be written as follows:
\begin{align}
F^+_{\mathrm{1}}(t,\theta_S,\phi_S,\psi_S) &= \frac{\sqrt{3}}{2}(\frac{1}{2}(1+\cos^2\theta_S)\cos2\phi_S(t)\cos2\psi_S \notag \\   &\quad-\cos\theta_S\sin2\phi_S(t)\sin2\psi_S)  \, , \\
F^\times_{\mathrm{1}}(t,\theta_S,\phi_S,\psi_S) &= \frac{\sqrt{3}}{2}(\frac{1}{2}(1+\cos^2\theta_S)\cos2\phi_S(t)\sin2\psi_S \notag \\   &\quad+\cos\theta_S\sin2\phi_S(t)\cos2\psi_S) \, , \\
F^+_{\mathrm{2}}(t,\theta_S,\phi_S,\psi_S)&= F^+_{\mathrm{1}}(t,\theta_S,\phi_S -\frac{\pi}{4},\psi_S) \, ,\\
F^\times_{\mathrm{2}}(t,\theta_S,\phi_S,\psi_S) &=F^\times_{\mathrm{1}}(t,\theta_S,\phi_S-\frac{\pi}{4},\psi_S)\, , 
\end{align}
where subscripts 1 and 2 are labels for the two Michelson signals, as indicated by the $\pi/4$ phase difference between the corresponding antenna pattern functions \cite{PhysRevD.57.7089}. 
Note that there is a coefficient $\sqrt{3}/2$ in response calculation when considering the $60^{\circ}$-interferometers as L-shaped interferometers.

\begin{table*}
    \caption{Key parameters for the configurations of TianQin, LISA, Taiji and DECIGO.}
    \centering
    \begin{tabular}{l | c | c | c | c}
		\hline
		\hline
		Parameters          &TianQin  &LISA  &Taiji  &DECIGO  \\
		\hline
		Number of satellites   & N=3     & N=3  &N=3    &N=$4\times3$=12     \\
		Orientation            & $\lambda=120.4^\circ$, $\beta=-4.7^\circ$   & cyclic & cyclic & cyclic\\
		Duty cycle    & 50\%  & 75\%  & 75\%  & 75\%  \\
		Mission lifetime (yr)      & 5   & 4  & 4  & 4 \\
		Arm length $L$(m)            & $\sqrt{3}\times10^8$  & $ 2.5\times10^9$ & $3\times10^9$  & $ 1\times10^6$ \\
		Displacement measurement noise $S_x$ (m$^2$Hz$^{-1}$)        & $ 1\times10^{-24}$  & $ 2.25\times10^{-22}$  & $ 6.4\times10^{-23}$ & $ 7.05\times10^{-36}$\\
		Acceleration noise $S_a$ (m$^2$s$^{-4}$Hz$^{-1}$)    & $ 1\times10^{-30}$  & $ 9\times10^{-30}$  & $ 9\times10^{-30}$  & $ 2.15\times10^{-37}$\\
        Radiation pressure noise $S_r$ (m$^2$s$^{-4}$Hz$^{-1}$) & -- & -- & -- & $7.48\times10^{-36}$  \\
		\hline
		\hline
	\end{tabular}
	\label{tb:parameter for detectors}
\end{table*}

\begin{table}
	\caption{The fitted coefficients of the galactic foreground spectra for Taiji}
	\begin{tabular}{c|c }
		\hline
		\hline
		Parameter & Taiji  \\
		\hline
		$a_0$       & -85.5448       \\
		$a_1$       & -3.23671       \\
		$a_2$       & -1.64187       \\
		$a_3$       & -1.14711       \\
		$a_4$       & 0.0325887      \\
		$a_5$       & 0.187854       \\
		\hline
		\hline
		\end{tabular}
	\label{tb:parameter for foreground}
\end{table}

\subsection{Sensitivity curve}
\label{sec:3c}

In \ac{GW} detection, a key performance metric of the detectors is their frequencies response to signal sources, which is characterized by the detector's sensitivity curve. 
This curve describes the detector's sensitivity across different frequencies and is closely related to the noise \ac{PSD}. 

The sensitivity curve of TianQin is given by \cite{2020PhRvD.102f3021H}
\begin{align}
S^{\text{TQ}}_n(f) &= \frac1{L^2}\left[\frac{4S_a}{(2\pi f)^4}\left(1+\frac{10^{-4}{\rm Hz}}{f}\right) +S_x\right] \notag \\ 
&\quad \times\left[1+0.6\left(\frac{f}{f_*}\right)\right],
\label{eq:S_N_TQ} 
\end{align}
where transfer frequency $f_*$ = $\frac{c}{2\pi L}$, arm length $L$, acceleration noise $S_a$, and displacement measurement noise $S_x$ are given in Table \ref{tb:parameter for detectors}. 

For both LISA and Taiji, they share an identical sensitivity curve of the form as \cite{2021arXiv210801167B,2025arXiv250416712H}
\begin{align}
S^{\text{LISA}}_n,S^{\text{Taiji}}_n(f) &= \frac1{L^2}\left[\frac{4S_a}{(2\pi f)^4}\left(1+\frac{4\times10^{-4}{\rm Hz}}{f}\right) \right. \notag \\
&\quad \left. \left(1+\left(\frac{f}{8\times10^{-3}}\right)^{4}\right)+S_x\right] \notag \\
&\quad \times\left[1+0.6\left(\frac{f}{f_*}\right)\right]+S_{conf}^{\text{LISA,Taiji}},
\label{eq:S_N_LISA} 
\end{align}
where $L$, $S_a$, $S_x$ are similarly given in Table \ref{tb:parameter for detectors}. 

While LISA and Taiji will resolve individual signals well, there are still a large number of faint signals from galactic \ac{DWD} systems that are not strong enough to be identified individually.
The \acp{GW} from these sources will form a significant foreground confusion noise for the sub-mHz observation, which can be characterized by its \ac{PSD} $S_{conf}$(f), being given by 
\begin{align}
S^{\text{LISA}}_{conf}(f) &= \frac{3}{10} \times 9 \times 10^{-45} \times f^{-\frac{7}{3}} \times \exp\left(-f^{0.138} \right. \notag  \\
&\quad \left. - 221 \times f \times \sin(521 \times f)\right) \times \left(1 + \tanh\left(1680 \right. \right. \notag \\ 
&\quad \left.\left. \times (0.00113 - f)\right)\right) ,
\label{eq:LISA foreground}
\end{align}
\begin{equation}
S^{\text{Taiji}}_{conf}(f) = \exp\left( \sum_{i=0}^{5} a_i \left( \ln \left( \frac{f}{m\rm Hz}\right)\right)^i\right),
\label{eq:Taiji foreground}
\end{equation}
where $a_i$ are the coefficients of polynomial, given in Table \ref{tb:parameter for foreground}.

For DECIGO, a dedicated analysis of the galactic confusion foreground and its effect on the detector's sensitivity curve has not yet been performed in the literature. We therefore adopt the instrumental noise only, as given by~\cite{2011PhRvD..83d4011Y}:
\begin{align}
S^{\text{DECIGO}}_n(f) &=  \frac{1}{L^2}\left[\frac{4S_a}{(2\pi f)^4}+S_x\left(1+\left(\frac{f}{f_p}\right)^{2}\right) \right. \notag \\
&\quad \left. +\frac{S_r}{(2\pi f)^4}\left(\frac{1}{1+\left(\frac{f}{f_p}\right)^{2}}\right)\right],
\label{eq:S_N_DECIGO} 
\end{align}
where $f_p$ = 7.36 Hz, $S_a$, $S_x$, radiation noise $S_r$ are given in Table \ref{tb:parameter for detectors}.

\begin{figure}
\includegraphics[width=0.5\textwidth]{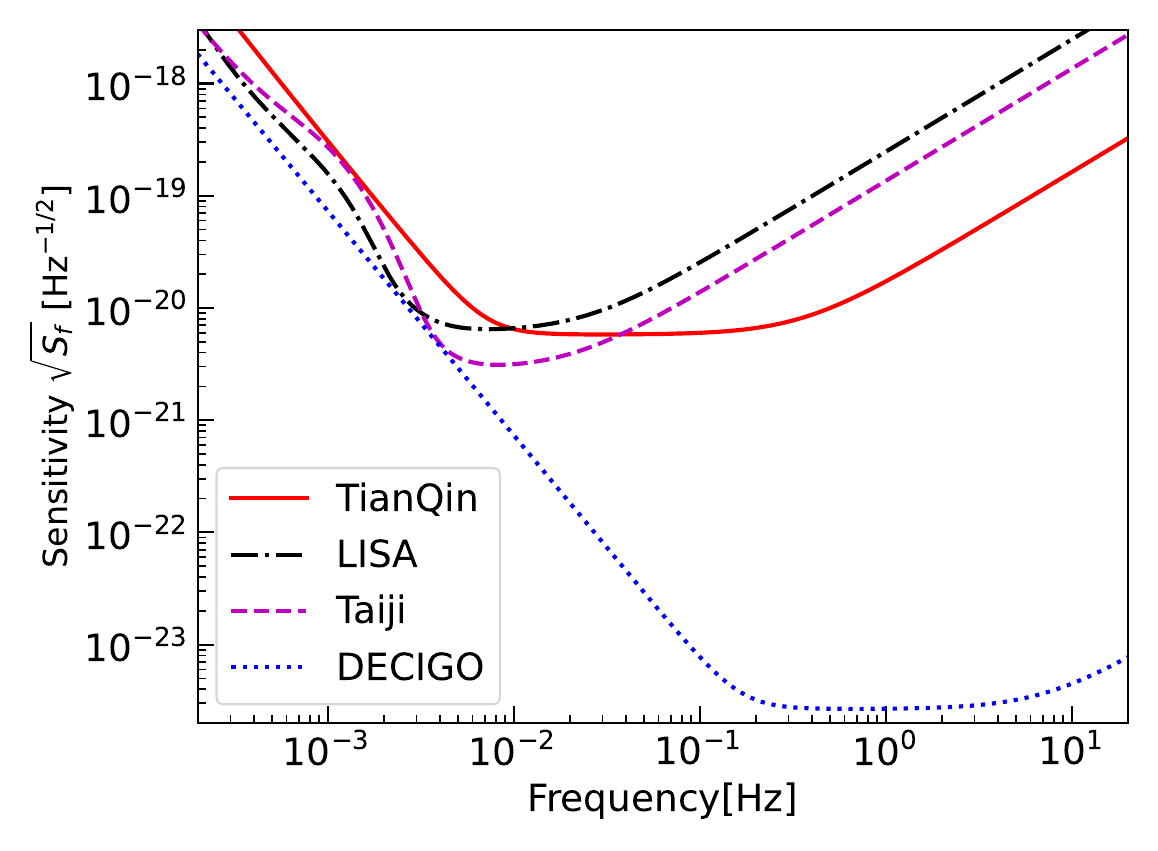}
\caption{The sensitivity curves of TianQin, LISA, Taiji and DECIGO. The red solid line, black dash-dot line, magenta dashed line, and blue dotted line correspond to TianQin, LISA, Taiji, and DECIGO respectively, defined in Eqs.~(\ref{eq:S_N_TQ})-(\ref{eq:S_N_DECIGO}). }
\label{fig:sensitivity}
\end{figure}

All the sensitivity curves illustrated above with their foreground confusion noise are represented in Fig. \ref{fig:sensitivity}.
This figure clearly delineates the frequency-dependent sensitivity of each detector, with DECIGO demonstrating the lowest noise floor across nearly the entire millihertz band, highlighting its superior design sensitivity. 
TianQin exhibits a comparable shape to LISA and Taiji but with a lower noise level above approximately 40 $m$Hz, consistent with its shorter arm length. 
All curves exhibit a characteristic rise at low frequencies due to acceleration noise, while at high frequencies they gradually increase as the transfer frequency term dominates.
Notably, the Galactic foreground confusion noise creates a “bump” in the LISA and Taiji curves below a few millihertz, representing the unresolved \acp{GW} from millions of \acp{DWD} in the Milky Way.

\subsection{Signal to noise ratio}
\label{sec:3d}

The \ac{SNR} $\rho$ of a signal is defined as
\begin{equation}
\rho^2 = (h|h),
\end{equation}
where the inner product $(\cdot|\cdot)$ is defined as
\begin{align} 
(a|b)&=4\Re e\int^\infty_0 \mathrm{d}f \frac{\tilde{a}^*(f)\tilde{b}(f)}{{S}_n(f)} \notag \\
&\quad\simeq\frac{2}{{S}_n(f)}\int^{T}_0 \mathrm{d}t \,a(t)b(t),
\label{eq:inner product}
\end{align}
where $\tilde{a}(f)$ and $\tilde{b}(f)$ are the Fourier transformations of two generic functions $a(t)$ and $b(t)$, ${S}_n(f)$ is the sensitivity curve of detector.

Therefore, the \ac{SNR} for a single binary can be calculated as
\begin{equation}
\rho^2 = (h|h) \simeq \frac{2}{{S}_n(f)}\int^{T}_0 \mathrm{d}t \,h(t)h(t) \,.
\label{eq:SNR}
\end{equation}
where $T$ is the observation time.

For a network of independent detectors, the total \ac{SNR} of a binary are given by:
\begin{equation}
\rho^2_{\mathrm{total}}=\sum_{a} \rho^2_a = \sum_{a} (h_a|h_a) \,,
\label{eq: tot_SNR}
\end{equation}
where the subscript $a$ stands for quantities related to the $a{\rm th}$
detector.

In addition, it is useful to define the characteristic strain $h_c = A \sqrt{N}$, with $N=fT$ being the number of binary orbital cycles observed during the mission.
Similarly, the noise characteristic strain is defined as
\begin{equation}
h_n(f)=\sqrt{f{S}_n(f)}\,,
\label{eq: h_n}
\end{equation}
one can straightforwardly estimate \ac{SNR} from the ratio between $h_c$ and $h_n$.

\subsection{Parameter estimation}
\label{sec:3e}

The uncertainty in the binary parameters can be derived
from the \ac{FIM} $\Gamma_{ij}$
\begin{equation}
\Gamma_{ij} = \left(\frac{\partial h}{\partial \xi_i}\bigg|\frac{\partial h}{\partial \xi_j}\right),
\label{eq:FIM}
\end{equation}
where $\xi_{i}$ stands for the $i{\rm th}$ parameter.

In the high-\ac{SNR} limit ($\rho\gg$1), the inverse matrix of the \ac{FIM} is equal to the covariance matrix, $\Sigma = \Gamma^{-1}$. 
The diagonal elements of the covariance matrix $\Sigma_{ii}$ give the variance of each parameter $\Delta\xi_{i}$, while the off-diagonal elements represent the covariance.

In numerical calculations, we approximate $\partial h /\partial \xi_i$ with numerical differentiation:
\begin{equation}
\frac{\partial h}{\partial\xi_i} \approx  \frac{\delta h}{\delta\xi_i}\equiv \frac{h(t,\xi_i+\delta\xi_i)-h(t,\xi_i-\delta\xi_i)}{2\delta\xi_i}\,,
\label{eq:partial}
\end{equation}
where the differentiation steps $\delta\xi_i$ should be chosen to make the variance $\Delta\xi_{i}$ stable \cite{2012A&A...544A.153S}.

Compared with the uncertainty of each coordinate, we are more interested in sky localization, which is a combination of the uncertainty of both coordinates \cite{1998PhRvD..57.7089C}:
\begin{equation}
\Delta\Omega_{\rm S} = 2\pi \big|\sin\beta \big| \big(\Sigma_{\beta\beta}\Sigma_{\lambda\lambda}-\Sigma^2_{\beta\lambda} \big)^{1/2}\,.
\label{eq:Omega}
\end{equation}

Similarly to \ac{SNR}, the total $\Gamma_{ij}$ of a network is given by:
\begin{equation}
\Gamma_{\mathrm{total}}=\sum_{a} \Gamma_a = \sum_{a} \left(\frac{\partial h_a}{\partial \xi_i}\bigg|\frac{\partial h_a}{\partial \xi_j}\right)\,,
\label{eq:tot_Gamma}
\end{equation} 
where the subscript $a$ stands for quantities related to the $a{\rm th}$
detector.

\section{result}\label{sec:4}

This section presents the results of our analysis of the four detectors mentioned earlier. 
We begin by evaluating the performance of each detector in standalone operation. 
We then explore the potential enhancement from operating multiple detectors simultaneously in a network configuration, such as TianQin + LISA, since coordinated observations can significantly improve \ac{GW} detection capabilities.
It should be noted that the designs and orbital configurations of these four detectors differ slightly, and their operation times also vary. 
Specifically, TianQin has a mission lifetime of 5 years with a duty cycle of 50\% \cite{2021PTEP.2021eA107M}, while LISA and Taiji share a mission lifetime of 4 years and a duty cycle of 75\% \cite{2022GReGr..54....3A,2025arXiv250416712H}. 
For DECIGO, we adopt the same mission lifetime and duty cycle as LISA and Taiji.
All relevant parameters for the four detectors are summarized in Table \ref{tb:parameter for detectors}.

In this work, we investigate several network configurations ranging from two to four detectors. 
Hereafter, the following notation is used: TL for TianQin + LISA, TD for TianQin + DECIGO, LD for LISA + DECIGO, TLD for TianQin + LISA + DECIGO, and TLTD for TianQin + LISA + Taiji + DECIGO.
In this study, binaries with \ac{SNR} $\rho \geq 5$ are treated as detectable sources. 
We examine the response of various detector configurations to the candidate \acp{VB} introduced in Section \ref{sec:2}, and further assess the uncertainties in parameter estimation for sources with $\rho \geq 20$.

\subsection{The SNR of verification binaries}\label{sec:4.1}

\begin{figure*}
\centering
\includegraphics[width=0.8\textwidth]{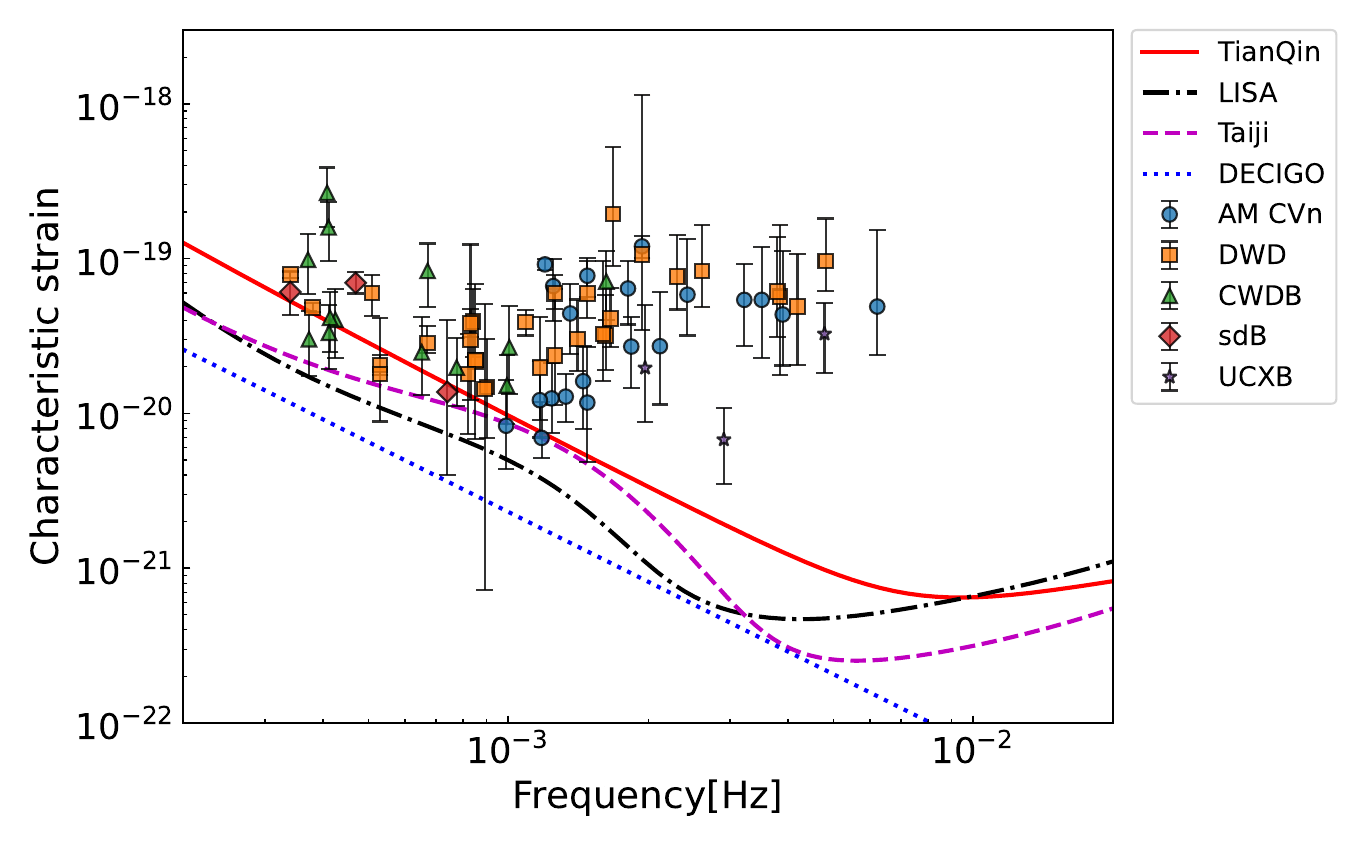}
\caption{Characteristic strain \(h_c\) versus frequency for the 72 verification 
binaries, compared with the detector noise characteristic strain curves 
\(h_n(f)\). 
The red solid, black dash-dot, magenta dashed, and blue dotted lines 
correspond to the sensitivity curves of TianQin, LISA, Taiji, and 
DECIGO, respectively, as given by Eqs.~\ref{eq:S_N_TQ} to 
\ref{eq:S_N_DECIGO}. 
Markers show \(h_c\) evaluated at the gravitational-wave frequency 
\(f\) using the measured parameter values from 
Table~\ref{tab:sample}: blue circles represent AM CVn systems, orange 
squares represent DWDs, green triangles represent CWDBs, red diamonds 
represent sdBs, and purple stars represent UCXBs.
Error bars indicate the range of \(h_c\) obtained by varying the 
binary parameters within their observational extremes, i.e., the 
maximum and minimum values taken from the error intervals in 
Table~\ref{tab:sample}.
}
\label{fig:strain}
\end{figure*}

We consider 72 sources, calculating the amplitude of each binary from the 
parameters listed in Table~\ref{tab:sample}. 
For each binary, the characteristic strain $h_c$ is evaluated at the 
gravitational-wave frequency $f$ using the measured values given in 
Table~\ref{tab:sample}, and the resulting points are shown as markers 
in Fig.~\ref{fig:strain}.

To indicate the possible range arising from observational uncertainties, 
we also compute $h_c$ using the maximum and minimum values of the 
parameters taken from the error intervals in Table~\ref{tab:sample}, 
or from typical uncertainties such as $\iota = 60^\circ \pm 10^\circ$ 
when individual error estimates are not available. 
The upper and lower bounds obtained in this way are displayed as error 
bars.

The characteristic strain of the binaries is displayed together with the 
detector noise characteristic strain curves \(h_n(f)\) in Fig.~\ref{fig:strain}. 
As shown in the figure, the sources are distributed throughout the millihertz band, with \acp{DWD} constituting the dominant population.
The characteristic strain primarily distributed between the orders of magnitude $10^{-20}$ and $10^{-19}$, with the maximum characteristic strain capable of reaching the magnitude of $10^{-18}$. 
While most sources reside above the DECIGO sensitivity curve (blue dotted line) and only 1 may lie below it, a significant portion fall below the curves of TianQin, LISA, and Taiji, especially for binaries with weak strain amplitudes.
This pattern indicates that DECIGO may offer the best individual detection performance, whereas the other three detectors have a more limited reach for the faintest sources.

In Table \ref{tb:VB_snr} (Appendix \ref{app:snr}), we report \acp{SNR} for all 72 sources under each of the nine considered detector configurations (namely, TianQin, LISA, Taiji, DECIGO, TL, TD, LD, TLD and TLTD), with settings $\phi_0=\pi$ and $\psi_s=\pi/2$.
Consistent with the sensitivity curves, DECIGO achieves the highest \acp{SNR}. 
For instance, sources like CP Eri, V406 Hya, and WD 1242 yield an \ac{SNR} less than 1 with TianQin, but more than 5 with DECIGO.
This means that DECIGO has the ability to detect more \acp{VB} that cannot be found by TianQin. 
The \acp{SNR} are obviously improved considering the network, as the \ac{SNR} of the network is given by the root sum squared of the \acp{SNR} of the individual detectors [see Eq. \eqref{eq: tot_SNR}]. 
For example, the SDSS J1908 source yields an \ac{SNR} of $\sim$ 9.2 with TianQin but reaches $\sim$ 93.8 with TLTD.

The overall detection efficiency across the source catalogue is quantified by counting sources above a range of \ac{SNR} thresholds. 
DECIGO alone can detect 71 of the 72 sources, or 98.6\% of the total, at a threshold of $\rho \geq 5$, substantially outperforming TianQin with 27 sources at 37.5\%, LISA with 42 sources at 58.3\%, and Taiji with 32 sources at 44.4\%.
The advantage of DECIGO persists even at higher \ac{SNR} thresholds: at $\rho \geq 10$, it detects 56 sources. 
Moreover, DECIGO remains the only single detector capable of identifying a significant number of high-\ac{SNR} systems, detecting 18 sources or 25.0\% of the total at $\rho \geq 100$.
Multi-detector networks significantly enhance these statistics. 
The complete TLTD configuration detects 56 sources, corresponding to 77.8\%, at $\rho \geq 10$.
Even at the stringent threshold of $\rho \geq 100$, networks that include DECIGO consistently identify at least 18 sources, representing 25.0\% of the catalogue. 
These counts quantitatively confirm that network operations increase the detectability of marginal sources and also expand the sample of high-confidence detections available for precision studies.

The distribution of extreme \ac{SNR} values across different detector configurations further illustrates the performance differences outlined earlier. 
Among the single-detector setups, TianQin yields \acp{SNR} from 0.376 for source WD 1704 to 116.137 for source J0806.
Notably, LISA and Taiji exhibit a complementary pattern: LISA provides a higher minimum \ac{SNR} of 1.185 from SDSS J0926 but a lower maximum of 124.373 from AM CVn, compared to Taiji’s range of 0.649 from SDSS J0926 to 240.838 from J0806. 
This is a direct consequence of their complementary frequency-dependent sensitivities.
LISA is more sensitive at lower frequencies, while Taiji performs better at higher frequencies. 
DECIGO, by contrast, covers a significantly broader dynamic range, from 4.852 for SDSS J0926 to 773.679 for J0806. 
In the complete TLTD configuration, the weakest source, SDSS J0926, reaches an \ac{SNR} of 5.083, while the strongest source, J0806, reaches 827.292.

This performance hierarchy is further quantified by the central tendencies. 
The mean and median \acp{SNR} are 9.859 and 2.328 for TianQin, 26.493 and 5.924 for LISA, 26.066 and 3.465 for Taiji, and 85.848 and 22.133 for DECIGO.
In particular, DECIGO demonstrates superior and more uniform sensitivity, achieving not only the highest maximum and a notably elevated minimum \ac{SNR}, but also the highest mean and median values, surpassing all other single detectors.
The synergy of the complete TLTD network further enhances these capabilities. 
The overall performance of the network, reflected in a mean \ac{SNR} of 95.071 and a median of 23.418, surpasses that of even the best single detector, DECIGO, conclusively underscoring the benefit of multi-detector observations.

\begin{figure*}
    \centering
    %

    \subfigure{
        \includegraphics[width=0.48\textwidth]{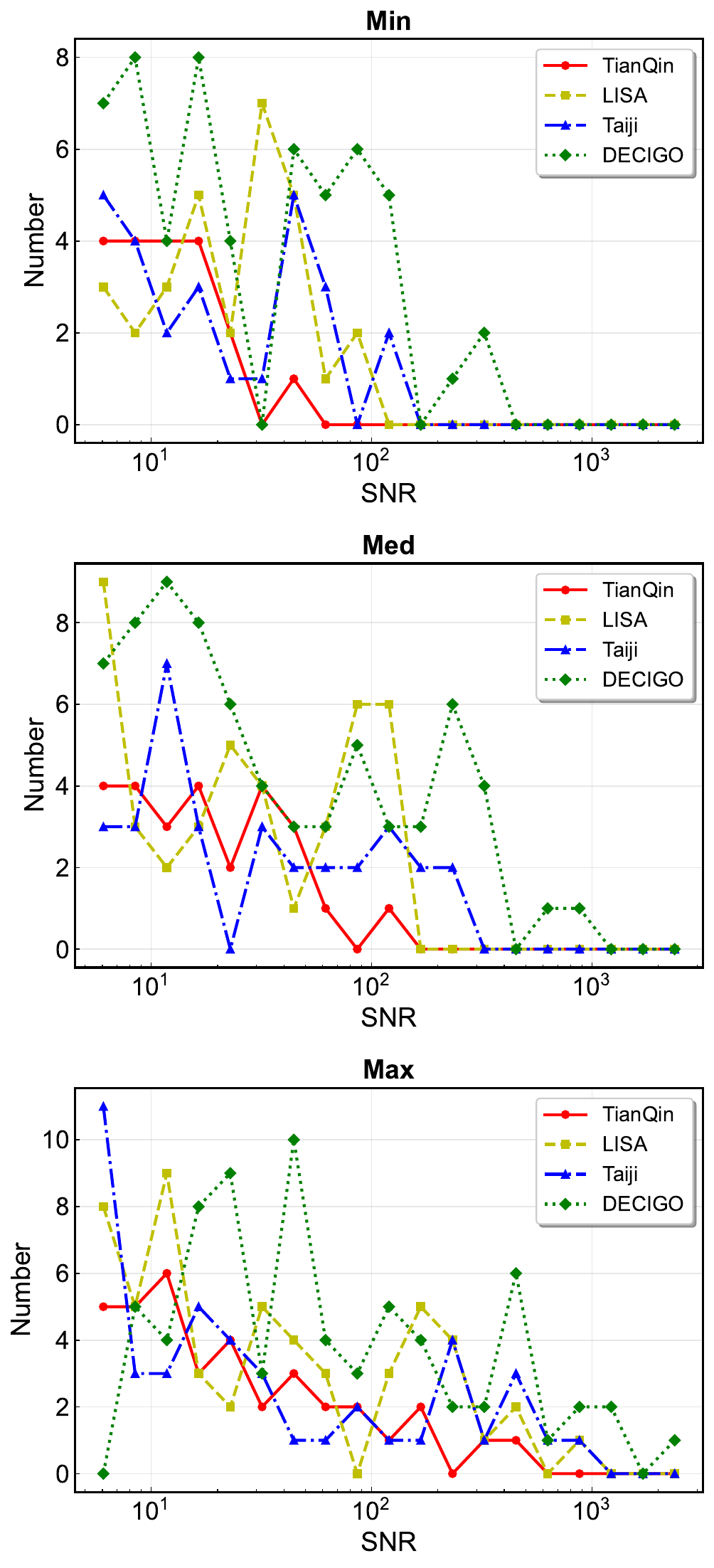}
        \label{fig:sub1}
    }
    \hfill
    \subfigure{
        \includegraphics[width=0.48\textwidth]{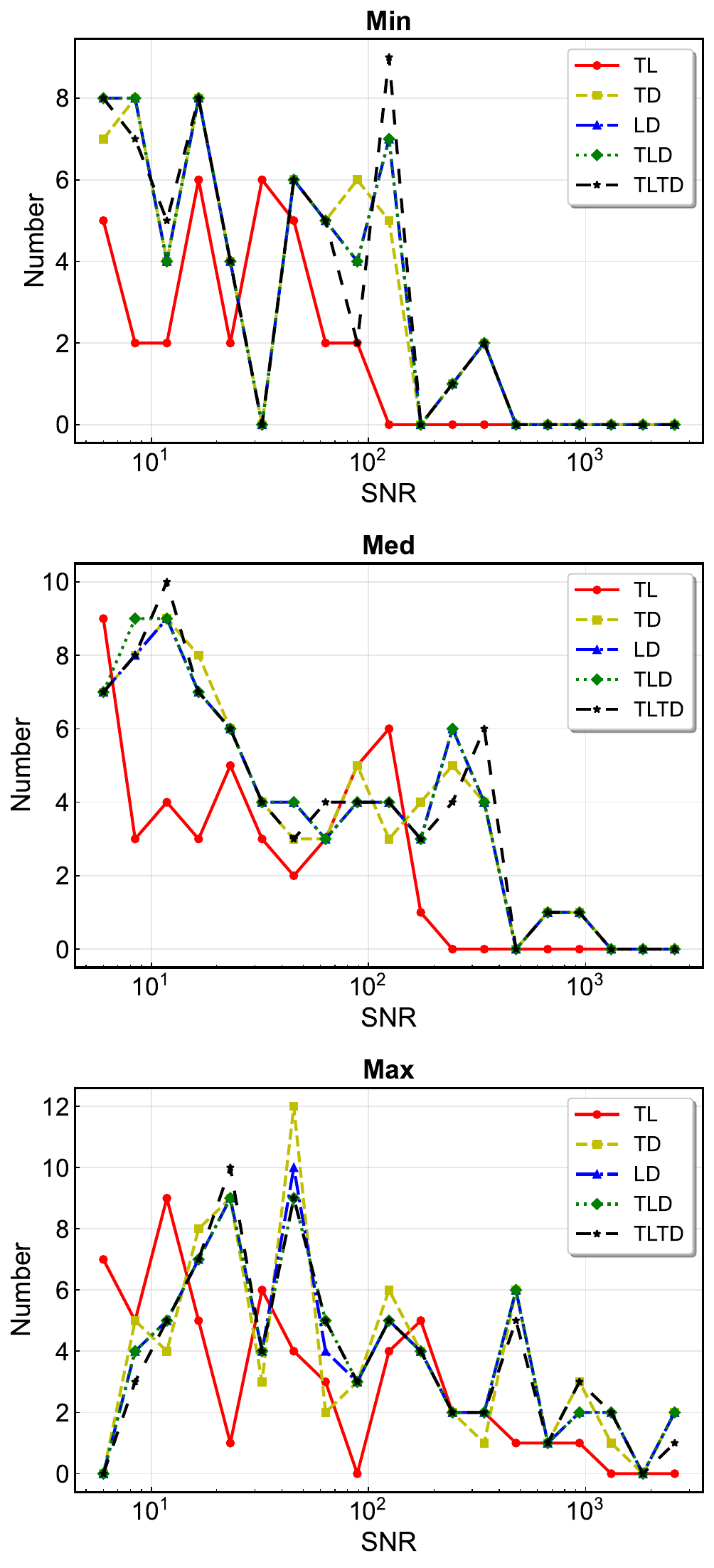}
        \label{fig:sub2}
    }
    
    \caption{The left column shows the results for single detectors (TianQin, LISA, Taiji, DECIGO), while the right column shows the results for multi-detector networks (TL, TD, LD, TLD, TLTD). For each configuration, the distributions of pessimistic (Min), most likely (Med), and optimistic (Max) estimates of the \ac{SNR} are provided. The underlying distributions are first constructed as histograms with 20 logarithmically spaced \ac{SNR} bins. To avoid visual clutter from heavily overlapping bars, only the bin centers and their corresponding counts are retained and connected by lines. The markers indicate the bin centers, and the vertical coordinate gives the number of sources in that bin.}

    \label{snr statistic}
\end{figure*}

\begin{figure*}
\centering
\includegraphics[width=0.8\textwidth]{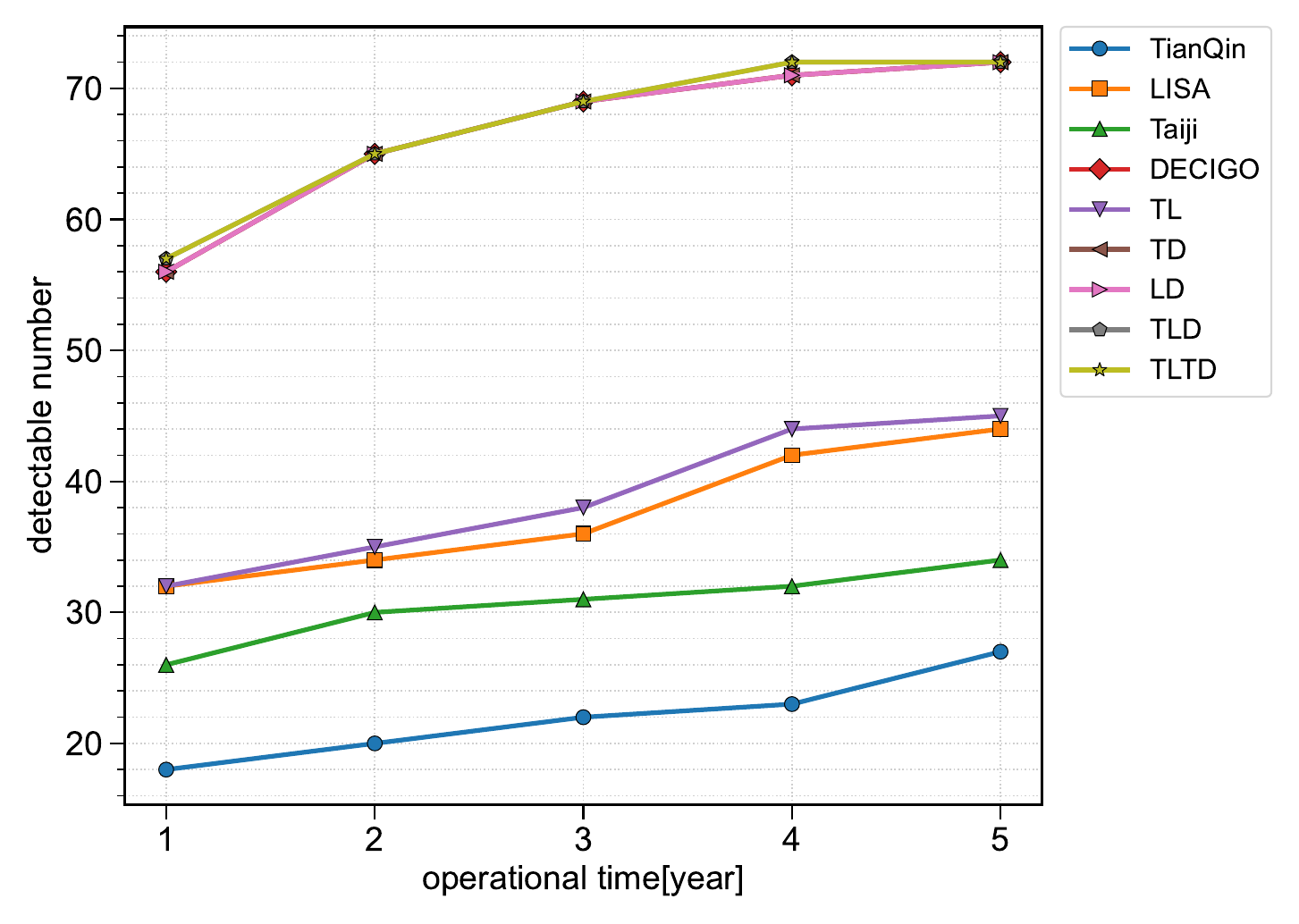}
\caption{The detectable number ($\rho \geq 5$) of the 72 \acp{VB} over time. Detectors are distinguished by both line colors and marker shapes: TianQin (blue circle), LISA (orange square), Taiji (green triangle up), DCIGO (red diamond), TL (purple triangle down), TD (brown triangle left), LD (pink triangle right), TLD (gray pentagon), and TLTD (olive star). }
\label{fig:observable numble}
\end{figure*}

In Fig.~\ref{snr statistic}, we show the distribution of the number of sources with \ac{SNR} greater than 5 in all configurations.
The underlying data are first constructed as a histogram.
For single-detector configurations, the range spans from a minimum \ac{SNR} just above 5 to the maximum \ac{SNR} of 2762.199 reached by DECIGO.
For multi-detector networks, the range spans from a minimum \ac{SNR} just above 5 to the maximum \ac{SNR} of 2997.777 reached by TLTD.
In both cases, the range is divided into 20 logarithmically spaced bins, and the number of sources in each bin is counted.
To avoid the visual clutter of heavily overlapping bars, the histogram is converted to a simplified display: only the bin center and its corresponding count are retained as a data point, and these points are connected by lines.

We also take into account the uncertainty of parameters: pessimistic (Min) and optimistic (Max) estimates correspond to the least and most favorable combinations of parameters, for example, maximum distance and minimum mass versus minimum distance and maximum mass, while the most likely estimate (Med) uses the median value of the parameters.
As anticipated, the curves shift systematically towards higher \ac{SNR} values from the Min to the Med to the Max scenario. 
The overall distributions are concentrated at lower \ac{SNR} values with an extended tail toward higher values, indicating that most systems achieve relatively low \ac{SNR}, with only a minority reaching high \ac{SNR} levels. 
A direct comparison of the curve heights across configurations reveals a clear hierarchy in the number of detectable sources. 
Among single-detector configurations, DECIGO, owing to its superior sensitivity in the millihertz band, shows both a higher total count and a distribution curve that extends further toward higher \ac{SNR} values, outperforming TianQin, LISA, and Taiji across most of the \ac{SNR} range. The introduction of multi-detector networks leads to a systematic upward shift of the curves across the entire \ac{SNR} range, with the distributions also extending further to the right. 
The complete TLTD network, in particular, results in the most substantial gain, significantly elevating the curve in the high-\ac{SNR} region.
This statistically confirms that coordinated multi-detector observations not only improve the overall \ac{SNR} of signals but, more importantly, transform a population of systems that would lie near the detection threshold in single-instrument mode into reliably detectable sources. 
Consequently, such networks effectively expand both the size and depth of the detectable sample, laying an observational foundation for a systematic \ac{GW} census of Galactic compact binary populations.

In addition, We calculated the \acp{SNR} of all sources as the operational time of different detectors varied from 1 to 5 years. 
Treating sources with $\rho \geq5$ as detectable sources, we calculated the variation over operational time in the number of detectable sources under different detectors. 
This result is shown in Figure \ref{fig:observable numble}.
The plot clearly reveals the substantial advantage of multi-detector networks and the critical role of long-duration observations.
Among single-detector configurations, DECIGO exhibits the fastest growth rate, requiring only about 1 year of observation to surpass the final 5-year detection count of other single instruments. 
This capability stems from its exceptional sensitivity in the millihertz range and the mode in which multiple satellites operate simultaneously (see Table.\ref{tb:parameter for detectors}).
TianQin shows a relatively flatter growth curve. 
This is influenced by its specific mission profile, which includes a 50\% duty cycle, while LISA and Taiji follow similar growth trajectories. 
The synergistic effect of multi-detector networks is particularly striking: the TLTD network maintains the highest or equal detectable-source count throughout, demonstrating the population-level gain from combining complementary detector sensitivities. 
This demonstrates that networks not only enhance the \ac{SNR} of individual sources through combined data but also systematically increase the total number of resolvable sources on a population level. 
This result strongly supports the scientific necessity of future coordinated multi-mission observations and maximizing mission lifetimes, which is essential for constructing a more complete census of Galactic compact binary \ac{GW} sources.

\subsection{the parameter estimation of verification binaries}\label{sec:4.2}

To assess the scientific potential of \ac{GW} observations, we performed a detailed parameter estimation for \acp{VB} with $\rho \geq$ 20 in various network configurations. 
Using the \ac{FIM} formalism described in Section \ref{sec:3e}, we quantified the measurement precision for key binary parameters including chirp mass ($\mathcal{M}_c$), distance ($d$), inclination angle ($\iota$) and sky localization ($\Omega$). 
Table \ref{tb:parameter_estimation1} (Appendix \ref{app:estimate}) illustrates the relative uncertainties for the 40 \acp{VB} under different detector configurations. 

Two types of blank entries appear in the table. 
If a source does not reach $\rho \geq 20$ in a particular detector configuration, all four parameter columns for that configuration are left blank. 
In a few cases the binary is detectable, but the estimated uncertainty in the inclination angle exceeds 90$^\circ$ and thus becomes physically unconstrained; for these systems, the $\iota$ entry is omitted while the uncertainties in chirp mass, distance, and sky localization are still reported.

As a single detector, DECIGO consistently delivers the highest parameter estimation precision due to its superior sensitivity in the millihertz band.
For high‑\ac{SNR} systems such as J0806 and ZTF J1539, DECIGO measures chirp masses with exquisite relative uncertainties below 0.2\%, constrains distances to within about 10\%, determines inclination angles $\iota$ to about 0.1\% for ZTF J1539 and 20\% for J0806, and achieves precise sky localization at the level of $\Delta\Omega_{\rm S} \sim 10^{-6} - 10^{-3} \, \text{deg}^2$. 
Nevertheless, it is noteworthy that although DECIGO provides the deepest reach, the average relative uncertainties across its entire catalogue are higher than those of other single detectors, with $\langle \sigma_{\mathcal{M}_c}/\mathcal{M}_c \rangle \sim 14.24$ and $\langle \sigma_d/d \rangle \sim 23.76$. 
This reflects its unique ability to probe fainter and more distant systems, which are inherently more difficult to characterize, thereby expanding the observable parameter space at the cost of larger average uncertainties for the faintest detections.

Multi-detector configurations significantly improve parameter estimation accuracy. 
For exceptionally high-\ac{SNR} systems such as J0806, the TLTD network achieves extraordinary precision: chirp mass relative uncertainty can drop to about 0.012\%, distance estimates are constrained to about 1.8\%, inclination angles to better than 5\%, and sky localization is refined to the level of $7\times10^{-5} \,\text{deg}^2$.
More importantly, this enhanced precision is not limited to a few outstanding sources but extends robustly across the entire detectable population. 
For the catalogue of 40 sources, the network yields dramatically reduced median uncertainties: $1.71 \times 10^{-4} \,\text{deg}^2$ for sky location, $4.21 \times 10^{-2} \,\text{M}_\odot$ for chirp mass, $0.340 \,\text{kpc}$ for distance, and $1.56^\circ$ for inclination. 

The chirp mass, which directly determines the GW amplitude, can be constrained with high precision for the majority of detectable sources. 
Among single detectors, Taiji achieves the tightest median constraint of $\sigma_{\mathcal{M}c} = 1.30 \times 10^{-2} \, \text{M}_{\odot}$ for its 16 sources with $\rho \geq 20$, followed by TianQin with $4.37 \times 10^{-2} \, \text{M}_{\odot}$ for 11 sources, and LISA with $8.91 \times 10^{-2} \, \text{M}_{\odot}$ for 24 sources. 
DECIGO, while detecting the largest sample of 39 sources, exhibits a significantly larger median uncertainty of $1.75 \times 10^{-1} \, \text{M}_{\odot}$ and the highest average relative uncertainty of $\langle \sigma_{\mathcal{M}_c}/\mathcal{M}_c \rangle \sim 14.24$, which reflects its unique sensitivity to a fainter and more challenging population of distant systems.
Network configurations dramatically improve and homogenize the measurement precision. 
The complete TLTD network delivers a median $\sigma_{\mathcal{M}c}$ of $4.21 \times 10^{-2} \, \text{M}_{\odot}$ across 40 sources --- a precision within a factor of about three of the best single detector, Taiji, but for a sample nearly three times as large. 
Furthermore, the average relative uncertainty of network drops to approximately 3.238, representing a several-fold improvement over DECIGO and clearly demonstrating the power of multi-detector networks to break parameter degeneracies.

Precise distance measurement is critical for determining the spatial positions of GW sources.
The performance of individual detectors varies significantly: Taiji again delivers the best median precision with $\sigma_d = 0.168 \, \text{kpc}$, while DECIGO exhibits a larger median uncertainty of $1.320 \, \text{kpc}$ and the highest average relative uncertainty, measured as $\langle \sigma_d/d \rangle \sim 23.76$. 
The synergy achieved in multi-detector networks substantially compresses these distance uncertainties.
For the complete TLTD network, the median $\sigma_d$ improves to $0.340 \, \text{kpc}$, and the average relative uncertainty decreases to approximately 5.418.
This uniform improvement across the entire catalogue, highlights the unique capability of network for distance determination.

The inclination angle, key to breaking the distance-inclination degeneracy, presents the widest range of uncertainties. 
Single detectors show significant scatter: Taiji and TianQin offer median precisions of $1.22^\circ$ and $3.43^\circ$, respectively.
DECIGO has a median uncertainty of $3.21^\circ$, but a subset of sources remains poorly constrained.
Networks improve the median inclination precision.
Even a two-detector network like TD reduces the median $\sigma_\iota$ to $1.70^\circ$. 
The complete TLTD network achieves a median of $1.56^\circ$. Nevertheless, a subset of sources remains poorly constrained in inclination, so reliable inclination determination is not available for the entire high-\ac{SNR} population.

Sky localization accuracy reflects the detector's ability to spatially locate a source. 
As a single detector, DECIGO achieves an excellent median \(\Delta\Omega_{\rm S}\) of \(4.5 \times 10^{-3}\,\text{deg}^2\). LISA and Taiji have larger medians of \(8.78 \times 10^{-2}\) and \(6.42 \times 10^{-2}\,\text{deg}^2\), respectively, whereas TianQin reaches a smaller median of \(3.21 \times 10^{-4}\,\text{deg}^2\).
We note that the absolute uncertainty in \(\Delta\Omega_{\rm S}\) is much smaller than the absolute uncertainties in chirp mass, distance, or inclination. 
There are two reasons for this.
First, the sky coordinates enter the waveform through the Doppler phase \(\Phi_D\) discussed in Sec.~\ref{sec:3a}, and the waveform is inherently more sensitive to phase than to amplitude.
Second, the phase accumulates a factor of order \(2\pi f t\) over the observation time \(t\), which can reach \(\sim 10^6\) for millihertz sources.
As a result, the constraints on \(\beta\) and \(\lambda\) become tighter 
the longer the source is observed.
In contrast, chirp mass, distance and inclination appear only in the amplitude and do not benefit from this temporal accumulation.
Here, the advantage of networks is most dramatic.
Each detector constrains the source direction from a different orbital position, and when these independent measurements are combined, the allowed sky region shrinks to their intersection.
The TL network reaches a median of \(3.76 \times 10^{-3}\,\text{deg}^2\), the TLD network improves to \(1.71 \times 10^{-4}\,\text{deg}^2\), and the complete TLTD network delivers the ultimate precision of \(1.71 \times 10^{-4}\,\text{deg}^2\).
This corresponds to localizing sources to areas on the sky thousands 
of times smaller than those achievable by most single detectors.

The results in Table~\ref{tb:parameter_estimation1} reveal that the 
relative performance of different detectors is not always uniform 
across parameters. 
For a given source, a detector may yield a tighter constraint on one 
quantity while giving a looser constraint on another.
This behaviour reflects the fact that different parameters are governed by different features of the waveform: the sky coordinates are constrained primarily by the Doppler phase, which depends on the 
detector's orbital motion, while the mass, distance, and inclination 
are inferred mainly from the signal amplitude and polarization, which are subject to different degeneracies.

A notable example is TianQin's performance on sources such as ZTF~J0546, ZTF~J1858, and SDSS~J0651. 
For these systems, TianQin yields a significantly smaller \(\Delta\Omega_{\rm S}\) than LISA, Taiji, and DECIGO, yet its constraints on \(\mathcal{M}_c\), \(d\), and \(\iota\) are comparatively weaker.
This can be traced to TianQin's geocentric orbit and fixed pointing, which produce a distinctive Doppler modulation pattern and a static 
antenna response on the sky.
For sources at favourable sky positions, the resulting phase modulation can be particularly well resolved, leading to a tight angular constraint even when the overall \ac{SNR} is only moderate.

\section{conclusion}\label{sec:5}

This study presents a systematic framework for evaluating the detectability and parameter estimation of Galactic binaries with the space-based \ac{GW} detectors TianQin, LISA, Taiji, and DECIGO. 
Moving beyond individual mission capabilities, we quantify the enhanced potential of a coordinated multi-detector network. 
Our analysis is grounded in a catalogue of known \acp{CVB} identified through \ac{EM} observations, and employs a combination of analytical expressions and numerical methods to compute instrument noise curves, \acp{SNR} and parameter estimation uncertainties.

Among the 72 \acp{VB} detectable by the TLTD network, individual detectors exhibit varying coverage: TianQin can detect 27, LISA 42, Taiji 32, and DECIGO 71 of these systems. 
The highest-\ac{SNR} source for TianQin, Taiji, and DECIGO is J0806, while for LISA it is AM CVn. Notably, DECIGO alone achieves the greatest single-detector \ac{SNR} for J0806, approaching 800. 
In all network configurations considered, J0806 remains the strongest source, with the complete TLTD network delivering the highest combined \ac{SNR} of over 800. 
This not only underscores the exceptional suitability of J0806 as a \ac{VB}, but also highlights the superior sensitivity of DECIGO to high-\ac{SNR} sources compared to other single detectors—an advantage that is further amplified in multi-detector networks.
Beyond the brightest sources, DECIGO yields higher \acp{SNR} than any other single detector across the entire catalogue, demonstrating its outstanding depth and breadth of detection, a capability that is enhanced even further in the combined TLTD configuration.

The distribution of detectable sources as a function of \ac{SNR} further elucidates the performance hierarchy.
Among single detectors, DECIGO not only yields the highest number of detections but also shifts the \ac{SNR} distribution markedly toward higher values, confirming its superior sensitivity within the millihertz band. 
The integration of detectors into networks, particularly the complete TLTD configuration, systematically increases the number of sources across all \ac{SNR} bins. 
Most notably, it transforms a substantial population of systems that would reside near the detection threshold for any single instrument into confidently detectable sources, thereby expanding the observable sample both in size and depth.

The number of detectable sources exhibits a strong dependence on the operational time. 
While all configurations show growth over time, DECIGO achieves the most rapid gain, surpassing the final 5-year detection count of other single detectors within approximately one year of observation. 
This rapid convergence underscores its exceptional design sensitivity. 
Multi-detector networks, however, unlock the most significant gains in survey efficiency.
The TLTD network consistently provides the highest or equal yield at any given mission time, while combining the complementary sensitivities of the four detectors and thereby providing a more comprehensive characterization of the source population than any individual detector.
This result underscores the critical importance of long, coordinated missions for compiling a complete census of Galactic \ac{GW} sources.

\ac{FIM} analysis reveals that key binary parameters-including chirp mass, distance, inclination, and sky location-can be constrained with remarkable precision for high-\ac{SNR} systems within a network.
While single detectors like Taiji provide excellent parameter estimation for the brightest sources, network configurations dramatically reduce uncertainties across the entire detectable population. 
The TLTD network, in particular, achieves median precisions of $\sim4.2\times10^{-2}\,\text{M}_{\odot}$ in chirp mass, $\sim$ 0.34 kpc in distance, $\sim1.56^\circ$ in inclination, and $\sim1.71\times10^{-4}\,\text{deg}^2$ in sky localization. 
This order-of-magnitude improvement in sky localization is pivotal, as it enables the unambiguous identification of \ac{EM} counterparts and facilitates detailed multi-messenger astrophysics.
Nevertheless, a full physical understanding of the detector-to-detector and parameter-to-parameter variations across the entire catalogue remains a complex task that goes beyond the scope of the present work. 
It involves the interplay of the source's frequency, sky position, and orientation with the orbit, pointing, and sensitivity of each detector. 
We defer a systematic investigation of these trends to a future study.

We are entering an era of multi-messenger synergy included \ac{EM} and \ac{GW} surveys.
We show that space-based \ac{GW} detectors, especially when operated as a coordinated network, have the potential to complement these surveys by delivering precise dynamical parameters for thousands of systems. 
This combined approach will enable population-level studies, constrain binary evolution models, and deepen our understanding of compact object physics across the Milky Way.

\section{Acknowledgments}
The work has been supported by the CPSF Postdoctoral Fellowship Program (Grant No. GZC20242112); Guangdong Provincial Key Laboratory of Advanced Particle Detection Technology (2024B1212010005); Guangdong Provincial Key Laboratory of Gamma-Gamma Collider and Its Comprehensive Applications (2024KSYS001); National Key Program for S\&T Research and Development (2023YFA1607200). 

\vspace{1cm}

\appendix
\section{table of verification binaries}\label{app:parameters}

\begin{widetext}
\setlength{\LTcapwidth}{\textwidth}
\begin{longtable}{l ccccccccc}        
  \caption{The Sample of Candidate Verification Binaries.
    Notes: 1. Square brackets represent the estimated values.
    2. The distance parameter is derived from the parallax provided by Gaia DR3.
    3. The right column uses Roman numerals to denote the sources from which the parameters of these sources are taken: (i) Ref. \cite{2018MNRAS.480..302K}, (ii) Ref. \cite{2010ApJ...711L.138R}, (iii) Ref. \cite{2026arXiv260107925C}, (iv) Ref. \cite{2023ApJ...953L...1B}, (v) Ref. \cite{2024ApJ...963..100K}, (vi) Ref. \cite{2017MNRAS.470.1894K}, (vii) Ref. \cite{2020ApJ...894L..15R}, (viii) Ref. \cite{2018A&A...620A.141R}, (ix) Ref. \cite{2025ApJ...987..205K}, (x) Ref. \cite{2011ApJ...739...34S}, (xi) Ref. \cite{2019Natur.571..528B}, (xii) Ref. \cite{2024ApJ...977..262C}, (xiii) Ref. \cite{2020ApJ...905L...7B}, (xiv) Ref. \cite{2020ApJ...905...32B}, (xv) Ref. \cite{2020ApJ...892L..35B}, (xvi) Ref. \cite{2019ApJ...886L..12B}, (xvii) Ref. \cite{2025ApJ...992..154M}, (xviii) Ref. \cite{2022ApJ...933...94B}, (xix) Ref. \cite{2025ApJ...991...65B}, (xx) Ref. \cite{2021ApJ...918L..14K}, (xxi) Ref. \cite{2023BSRSL..9211283S}, (xxii) Ref. \cite{2020ApJ...904...56G}, (xxiii) Ref. \cite{2023ApJ...950..141K}, (xxiv) Ref. \cite{2020MNRAS.494L..91C}, (xxv) Ref. \cite{2021ApJ...921..160C}, (xxvi) Ref. \cite{2000MNRAS.314..334M}, (xxvii) Ref. \cite{2023ApJS..264...39R}, (xxviii) Ref. \cite{2017ApJ...851...28K}, (xxix) Ref. \cite{2024MNRAS.527.2072D}, (xxx) Ref. \cite{2005A&A...440..287I}, (xxxi) Ref. \cite{2015ApJ...798..117P}, and the references therein.
    }
  \label{tab:sample}\\
  \hline
  \hline
    Source  & {$\lambda$} & {$\beta$} & {$f$} & {$d$} & {$M$} & {$m$} & {$\iota$} & Refs. \\
            & {[\si{deg}]} & {[\si{deg}]} & {[\si{mHz}]} & {[\si{kpc}]} & {[$\text{M}_{\odot}$]} & {[$\text{M}_{\odot}$]} & {[\si{deg}]} & \\
    \hline
    \hline
    \endfirsthead
  \multicolumn{9}{c}%
  {\tablename\ \thetable\ -- \textit{Continued from previous page}} \\
  \hline
  \hline
    Source  & {$\lambda$} & {$\beta$} & {$f$} & {$d$} & {$M$} & {$m$} & {$i$} & Refs. \\
            & {[\si{deg}]} & {[\si{deg}]} & {[\si{mHz}]} & {[\si{kpc}]} & {[$\text{M}_{\odot}$]} & {[$\text{M}_{\odot}$]} & {[\si{deg}]} & \\
    \hline
    \hline
  \endhead
  
  \hline
  \multicolumn{9}{r}{\textit{Continued on next page}} \\
  \endfoot
  
  \endlastfoot
    \multicolumn{10}{c}{\textbf{\ac{AM CVn}}}\\
J0806  & 120.4425 & -4.704 & 6.22 & $[5.0 \pm 3.0]$ & $0.55 \pm 0.05$ & $0.27 \pm 0.05$ & 38 & i,ii \\
ATLAS J1013  &179.4384 & -51.0003 & 3.89 & $1.802 \pm 0.639$ & $0.87^{+0.36}_{-0.25}$ & $0.10^{+0.03}_{-0.02}$ & [82] & iii \\
V407 Vul  & 294.9945 & 46.7829 & 3.51 & $2.087 \pm 0.676$ & $[0.8 \pm 0.1]$ & $[0.177 \pm 0.071]$ & [60] & i \\
ES Cet  & 24.612 & -20.3339 & 3.22 & $1.727 \pm 0.21$ & $[0.8 \pm 0.1]$ & $[0.161 \pm 0.064]$ & [60] & i \\
ZTF J0127  & 43.1707 & 40.009 & 2.43 & $1.283 \pm 0.604$ & $0.75 \pm 0.06$ & $0.19 \pm 0.03$ & [80] & iv \\
SDSS J1351  & 208.3879 & 4.4721 & 2.12 & $1.341 \pm 0.436$ & $[0.8 \pm 0.1]$ & $[0.1 \pm 0.040]$ & [60] & i \\
AM CVn  & 170.3858 & 37.4427 & 1.94 & $0.302 \pm 0.003$ & $0.68 \pm 0.06$ & $0.125 \pm 0.012$ & 43 & i \\
SDSS J1908  & 298.2172 & 61.4542 & 1.84 & $0.977 \pm 0.032$ & $[0.8 \pm 0.1]$ & $[0.085 \pm 0.034]$ & 15 & i \\
HP Lib  & 235.0882 & 4.9597 & 1.81 & $0.28 \pm 0.002$ & $0.645 \pm 0.155$ & $0.068 \pm 0.02$ & 30 & i \\
TIC378898110  & 215.5392 & -52.6045 & 1.48 & $0.309 \pm 0.002$ & $[0.8 \pm 0.1]$ & $0.1^{+0.02}_{-0.02}$ & $74 \pm 10.0$ & v \\
PTF1 J1919  & 309.0023 & 69.029 & 1.48 & $1.365 \pm 0.474$ & $[0.8 \pm 0.1]$ & $[0.066 \pm 0.026]$ & [60] & i \\
CXOGBS J1751  & 268.0614 & -6.2526 & 1.45 & $0.939 \pm 0.102$ & [$0.8 \pm 0.1]$ & $[0.064 \pm 0.026]$ & [60] & i \\
CR Boo  & 202.2728 & 17.8971 & 1.36 & $0.352 \pm 0.005$ & $0.885 \pm 0.215$ & $0.066 \pm 0.022$ & 30 & i \\
KL Dra  & 334.1334 & 78.3217 & 1.33 & $0.922 \pm 0.088$ & $0.76 \pm 0.1$ & $0.057 \pm 0.01$ & [60] & vi,vii \\
V803 Cen  & 216.1673 & -30.3166 & 1.25 & $0.287 \pm 0.005$ & $0.975 \pm 0.195$ & $0.084 \pm 0.025$ & 13.5 & i \\
PTF1 J0719  & 104.3883 & 26.5213 & 1.24 & $0.841 \pm 0.201$ & $[0.8 \pm 0.1]$ & $[0.053 \pm 0.01]$ & [60] & viii \\
SMSS J1138  & 202.6662 & -47.997 & 1.2 & $0.547 \pm 0.018$ & $0.99^{+0.01}_{-0.01}$ & $0.24^{+0.01}_{-0.01}$ & $88.7 \pm 0.1$ & ix \\
SDSS J0926  & 132.2867 & 20.2342 & 1.18 & $0.814 \pm 0.165$ & $0.84 \pm 0.05$ & $0.029^{+0.009}_{-0.002}$ & 82.6 & x \\
CP Eri  & 42.1327 & -26.4276 & 1.17 & $0.748 \pm 0.203$ & $[0.8 \pm 0.1]$ & $[0.049 \pm 0.01]$ & [60] & i,viii \\
V406 Hya  & 140.7336 & -21.2342 & 0.99 & $0.736 \pm 0.231$ & $[0.8 \pm 0.1]$ & $[0.04 \pm 0.01]$ & [60] & i,viii \\

    \multicolumn{10}{c}{\textbf{\ac{DWD}}}\\
ZTF J1539  & 205.0315 & 66.1616 & 4.82 & $1.629 \pm 0.693$ & $0.61^{+0.017}_{-0.022}$ & $0.21^{+0.014}_{-0.015}$ & 84 & xi \\
ZTF J0546  & 87.2578 & 15.3132 & 4.19 & $3.707^{+1.631}_{-1.258}$ & $[0.6 \pm 0.1]$ & $[0.3 \pm 0.1]$ & [60] & xii \\
ZTF J1858  & 283.1123 & 2.3905 & 3.84 & $2.895^{+2.733}_{-1.449}$ & $[0.6 \pm 0.1]$ & $[0.3 \pm 0.1]$ & [60] & xii \\
ZTF J2243  & 13.2423 & 53.9599 & 3.79 & $1.753 \pm 0.72$ & $0.323 \pm 0.065$ & $0.335 \pm 0.054$ & $82.1 \pm 0.5$ & xiii \\
SDSS J0651  & 101.3396 & 5.8048 & 2.61 & $0.958 \pm 0.371$ & $0.50 \pm 0.04$ & $0.26 \pm 0.04$ & 86.9 & i \\
ZTF J0538  & 84.8061 & -3.4356 & 2.31 & $0.997 \pm 0.364$ & $0.45 \pm 0.05$ & $0.32 \pm 0.03$ & $85.4 \pm 0.1$ & xiv \\
ZTF J1905  & 293.7825 & 53.6335 & 1.94 & $0.696 \pm 0.602$ & $[0.6 \pm 0.1]$ & $[0.3 \pm 0.1]$ & [60] & xiv \\
SDSS J0935  & 130.9795 & 28.0912 & 1.68 & $0.396 \pm 0.205$ & $0.312 \pm 0.019$ & $0.75 \pm 0.24$ & [60] & i \\
SDSS J2322  & 353.4373 & 8.4572 & 1.66 & $0.86 \pm 0.204$ & $0.31 \pm 0.02$ & $0.29 \pm 0.05$ & 27 & xv\\
PTF J0533  & 82.9097 & -21.1234 & 1.62 & $1.172 \pm 0.389$ & $0.652 \pm 0.04$ & $0.167 \pm 0.03$ & 72.8 & xvi \\
ZTF J2029  & 314.4631 & 33.4205 & 1.6 & $1.095 \pm 0.643$ & $0.32 \pm 0.04$ & $0.3 \pm 0.04$ & $86.6 \pm 0.1$ & xvii \\
J1239  & 197.4157 & -15.0625 & 1.48 & $0.972 \pm 0.272$ & $0.291 \pm 0.013$ & $0.68^{+0.11}_{-0.06}$ & $71^{+8.0}_{-10.0}$ & v, xviii \\
ZTF J0722  & 115.8782 & -40.2164 & 1.41 & $1.267 \pm 0.442$ & $0.38 \pm 0.04$ & $0.33 \pm 0.03$ & $89.7 \pm 0.2$ & xvii \\
ZTF J1749  & 267.0209 & 32.8576 & 1.26 & $1.291 \pm 0.665$ & $0.4 \pm 0.07$ & $0.28 \pm 0.05$ & $85.5 \pm 1.4$ & xiv \\
SDSS J0634  & 97.0832 & 14.839 & 1.26 & $0.435^{+0.016}_{-0.015}$ & $0.45^{+0.07}_{-0.06}$ & $0.21^{+0.03}_{-0.02}$ & $43^{+7.0}_{-5.6}$ & xix \\
ZTF J2228  & 7.0683 & 53.2065 & 1.17 & $2.076 \pm 0.657$ & $[0.6 \pm 0.1]$ & $[0.3 \pm 0.1]$ & [60] & xiv \\
SMSS J0338  & 286.4357 & -72.7068 & 1.09 & $0.536 \pm 0.016$ & $0.23 \pm 0.015$ & $0.38 \pm 0.05$ & $69 \pm 9.0$ & xx \\
ZTF J0643  & 101.5702 & -19.6948 & 0.9 & $2.04 \pm 0.597$ & $[0.6 \pm 0.1]$ & $[0.3 \pm 0.1]$ & [60] & xxi \\
ZTF J0640  & 99.6432 & -5.4567 & 0.89 & $1.576 \pm 0.62$ & $0.39 \pm 0.12$ & $0.325 \pm 0.3$ & $65.3 \pm 5.1$ & xiv \\
SDSS J0106  & 11.4582 & -15.7928 & 0.85 & $0.825 \pm 0.438$ & $0.188 \pm 0.1$ & $0.57 \pm 0.01$ & 67 & i,vi \\
ZTF J2130  & 348.1685 & 54.4443 & 0.85 & $1.307 \pm 0.043$ & $0.545 \pm 0.02$ & $0.337 \pm 0.015$ & $86.4 \pm 1.0$ & xxii \\
SDSS J1630  & 231.7612 & 63.0501 & 0.84 & $0.848 \pm 0.167$ & $0.298 \pm 0.019$ & $0.76 \pm 0.24$ & [60] & i \\
SDSS J0822  & 120.6816 & 11.0965 & 0.83 & $0.837 \pm 0.574$ & $0.304 \pm 0.1$ & $0.524 \pm 0.01$ & 88.1 & i,vi \\
J1526  & 235.9907 & -8.1906 & 0.83 & $0.625 \pm 0.075$ & $0.37 \pm 0.02$ & $0.4 \pm 0.02$ & [60] & xxiii \\
ZTF J1901  & 306.8131 & 74.6335 & 0.82 & $0.909 \pm 0.078$ & $0.5 \pm 0.1$ & $0.2 \pm 0.1$ & 86.2 & xxiv \\
SDSS J1235  & 181.7907 & 17.9525 & 0.67 & $0.446 \pm 0.028$ & $0.35 \pm 0.01$ & $0.27^{+0.06}_{-0.02}$ & $27 \pm 3.8$ & v \\
SDSS J0056  & 10.6273 & -11.3044 & 0.53 & $0.627 \pm 0.046$ & $0.18 \pm 0.01$ & $0.82 \pm 0.02$ & [60] & i,vi \\
SDSS J1056  & 130.4076 & 52.2268 & 0.53 & $1.198 \pm 0.536$ & $0.334 \pm 0.1$ & $0.76 \pm 0.01$ & [60] & i,vi \\
SDSS J0923  & 133.7151 & 14.4268 & 0.51 & $0.288 \pm 0.005$ & $0.275 \pm 0.015$ & $0.76 \pm 0.23$ & [60] & i \\
WD 0957  & 208.5263 & -67.3013 & 0.38 & $0.164 \pm 0.001$ & $0.37 \pm 0.013$ & $0.32 \pm 0.013$ & 68 & i,vi \\
SDSS J1337  & 182.8931 & 45.5716 & 0.34 & $0.114 \pm 0.001$ & $0.51 \pm 0.01$ & $0.32 \pm 0.01$ & $34 \pm 1.0$ & xxv \\
WD 1242  & 194.5586 & -5.552 & 0.19 & $0.04 \pm 0.0$ & $0.56^{+0.05}_{-0.07}$ & $0.39^{+0.04}_{-0.05}$ & 45.1 & i,vi \\
WD 1704  & 242.3234 & 70.1865 & 0.16 & $0.039 \pm 0.0$ & $0.39 \pm 0.05$ & $0.56 \pm 0.07$ & [60] & xxvi \\

    \multicolumn{10}{c}{\textbf{\ac{CWDB}}}\\
ZTF J0526  & 84.6989 & 36.2955 & 1.63 & $0.845 \pm 0.067$ & $[0.6 \pm 0.1]$ & $[0.3 \pm 0.1]$ & $[60]$ & xxvii \\
ZTF J1840  & 279.7729 & 5.3797 & 1.01 & $1.286 \pm 0.277$ & $[0.6 \pm 0.1]$ & $[0.3 \pm 0.1]$ & $[60]$ & xxvii \\
ZTF J1707  & 257.2883 & 7.4854 & 0.99 & $2.236 \pm 0.176$ & $[0.6 \pm 0.1]$ & $[0.3 \pm 0.1]$ & $[60]$ & xxvii \\
ZTF J2351  & 36.8445 & 55.5895 & 0.78 & $1.28 \pm 0.081$ & $[0.6 \pm 0.1]$ & $[0.3 \pm 0.1]$ & $[60]$ & xxvii \\
ZTF J1611  & 183.9170 & 78.0916 & 0.67 & $0.257 \pm 0.008$ & $[0.6 \pm 0.1]$ & $[0.3 \pm 0.1]$ & $[60]$ & xxvii \\
ZTF J1813  & 276.0031 & 66.2346 & 0.65 & $0.834 \pm 0.12$ & $[0.6 \pm 0.1]$ & $[0.3 \pm 0.1]$ & $[60]$ & xxvii \\
ZTF J1603  & 232.8645 & 41.6031 & 0.43 & $0.312 \pm 0.023$ & $[0.6 \pm 0.1]$ & $[0.3 \pm 0.1]$ & $[60]$ & xxvii \\
ZTF J0451  & 71.5210 & -21.2702 & 0.41 & $0.294 \pm 0.003$ & $[0.6 \pm 0.1]$ & $[0.3 \pm 0.1]$ & $[60]$ & xxvii \\
ZTF J0112  & 44.5245 & 45.8154 & 0.41 & $0.365 \pm 0.013$ & $[0.6 \pm 0.1]$ & $[0.3 \pm 0.1]$ & $[60]$ & xxvii \\
ZTF J2334  & 12.6230 & 38.0721 & 0.41 & $0.076 \pm 0.0$ & $[0.6 \pm 0.1]$ & $[0.3 \pm 0.1]$ & $[60]$ & xxvii \\
ZTF J2007  & 309.0298 & 36.9204 & 0.41 & $0.045 \pm 0.0$ & $[0.6 \pm 0.1]$ & $[0.3 \pm 0.1]$ & $[60]$ & xxvii \\
ZTF J1943  & 293.0642 & -5.5741 & 0.37 & $0.358 \pm 0.015$ & $[0.6 \pm 0.1]$ & $[0.3 \pm 0.1]$ & $[60]$ & xxvii \\
ZTF J1708  & 258.3723 & -2.8898 & 0.37 & $0.109 \pm 0.001$ & $[0.6 \pm 0.1]$ & $[0.3 \pm 0.1]$ & $[60]$ & xxvii \\

    \multicolumn{10}{c}{\textbf{\ac{sdB}}}\\
OW J0741  & 125.3546 & -50.1862 & 0.74 & $1.576 \pm 0.644$ & $0.23 \pm 0.12$ & $0.72 \pm 0.17$ & 57.4 & xxviii \\
CD-30  & 221.1718 & -16.6125 & 0.47 & $0.355 \pm 0.006$ & $0.47^{+0.07}_{-0.06}$ & $0.74 \pm 0.02$ & 82.9 & xxix \\
HD 265435  & 101.3387 & 10.1442 & 0.34 & $0.461 \pm 0.012$ & $0.63^{+0.13}_{-0.12}$ & $1.01 \pm 0.15$ & $64^{+14.0}_{-5.0}$ & v \\

    \multicolumn{10}{c}{\textbf{\ac{UCXB}}}\\
1RXS J1718  & 261.7319 & -17.35 & 4.79 & $6.5 \pm 0.5$ & $[0.6 \pm 0.1]$ & $[0.3 \pm 0.1]$ & $[60]$ & xxx \\
4U 1820–30  & 275.1436 & -7.0275 & 2.91 & $7.972 \pm 0.277$ & $[1.4 \pm 0.6]$ & $[0.069 \pm 0.015]$ & [60] & v \\
4U 0513-40  & 70.602 & -62.7122 & 1.97 & $6.466^{+4.041}_{-3.134}$ & $[0.6 \pm 0.1]$ & $[0.55 \pm 0.1]$ & $[60]$ & xxxi \\
    \hline

\end{longtable}

\end{widetext}

\vspace{1cm}

\section{snr of verification binaries}\label{app:snr}

The \ac{SNR} of all selected binaries are listed in Table \ref{tb:VB_snr}, assuming a nominal mission lifetime of five years for TianQin and four years for LISA, Taiji, and DECIGO, with settings $\phi_0=\pi$ and $\psi_S=\pi/2$ for all binaries.

\begin{widetext}
\setlength{\LTcapwidth}{\textwidth}
\begin{longtable}{ccccccccccc}
\caption{The \ac{SNR} of \acp{VB} with single detector and the network of them. $\mathcal{A}$ is given in units in $10^{-23}$.} 
\label{tb:VB_snr}\\
\hline
\hline
\multicolumn{2}{c}{} & \multicolumn{9}{c}{\ac{SNR}} \\ 
\cline{3-11} 
\multicolumn{1}{c}{Source} & \multicolumn{1}{c}{$\mathcal{A}$} &
\multicolumn{1}{c}{TQ} & \multicolumn{1}{c}{LISA} & 
\multicolumn{1}{c}{Taiji} & \multicolumn{1}{c}{DECIGO} & \multicolumn{1}{c}{TL} & \multicolumn{1}{c}{TD} & \multicolumn{1}{c}{LD} & \multicolumn{1}{c}{TLD} & \multicolumn{1}{c}{TLTD} \\
\hline
\hline
\endfirsthead
\multicolumn{10}{c}%
{\tablename\ \thetable\ -- \textit{Continued from previous page}} \\
\hline
\hline
\multicolumn{2}{c}{} & \multicolumn{9}{c}{SNR} \\ 
\cline{3-11} 
\multicolumn{1}{c}{Source} & \multicolumn{1}{c}{$\mathcal{A}$} &
\multicolumn{1}{c}{TQ} & \multicolumn{1}{c}{LISA} & 
\multicolumn{1}{c}{Taiji} & \multicolumn{1}{c}{DECIGO} & \multicolumn{1}{c}{TL} & \multicolumn{1}{c}{TD} & \multicolumn{1}{c}{LD} & \multicolumn{1}{c}{TLD} & \multicolumn{1}{c}{TLTD}\\
\hline
\hline
\endhead
\hline \multicolumn{10}{r}{\textit{Continued on next page}} \\
\endfoot
\hline
\endlastfoot
\multicolumn{11}{c}{\textbf{\ac{AM CVn}}}\\
J0806                    &   6.39 &  116.137 &  119.761 &  240.838 &  773.679 &  166.824 &  782.347 &  782.893 &  791.460 &  827.292 \\
ATLAS J1013              &   7.18 &   16.159 &   57.384 &   84.306 &  163.386 &   59.615 &  164.183 &  173.170 &  173.922 &  193.278 \\
V407 Vul                 &   9.40 &   35.518 &   97.390 &  119.243 &  259.047 &  103.665 &  261.470 &  276.749 &  279.019 &  303.431 \\
ES Cet                   &   9.81 &   16.294 &   97.787 &   99.168 &  232.607 &   99.135 &  233.177 &  252.325 &  252.851 &  271.602 \\
ZTF J0127                &  12.20 &    9.652 &   50.395 &   27.919 &  106.841 &   51.312 &  107.276 &  118.130 &  118.523 &  121.767 \\
SDSS J1351               &   6.07 &    4.372 &   26.341 &   12.568 &   63.980 &   26.701 &   64.129 &   69.190 &   69.328 &   70.458 \\
AM CVn                   &  28.01 &   30.917 &  124.373 &   57.404 &  309.407 &  128.158 &  310.948 &  333.468 &  334.898 &  339.782 \\
SDSS J1908               &   6.48 &    9.209 &   28.779 &   13.269 &   87.797 &   30.217 &   88.279 &   92.394 &   92.851 &   93.795 \\
HP Lib                   &  15.50 &   16.372 &   62.874 &   29.044 &  172.165 &   64.970 &  172.941 &  183.286 &  184.016 &  186.294 \\
TIC378898110             &  20.72 &    6.211 &   22.022 &   10.916 &   79.333 &   22.882 &   79.576 &   82.333 &   82.567 &   83.285 \\
PTF1 J1919               &   3.14 &    1.495 &    4.142 &    2.053 &   16.878 &    4.403 &   16.944 &   17.379 &   17.443 &   17.563 \\
CXOGBS J1751             &   4.36 &    3.124 &    6.620 &    3.312 &   21.498 &    7.320 &   21.723 &   22.494 &   22.710 &   22.950 \\
CR Boo                   &  12.33 &    5.237 &   21.622 &   11.138 &   77.095 &   22.247 &   77.272 &   80.069 &   80.240 &   81.010 \\
KL Dra                   &   3.62 &    1.149 &    3.480 &    1.810 &   15.827 &    3.664 &   15.868 &   16.205 &   16.245 &   16.346 \\
V803 Cen                 &  19.33 &    7.476 &   29.203 &   15.607 &  114.214 &   30.145 &  114.458 &  117.888 &  118.125 &  119.151 \\
PTF1 J0719               &   3.65 &    1.887 &    3.312 &    1.776 &   12.660 &    3.811 &   12.800 &   13.086 &   13.222 &   13.340 \\
SMSS J1138               &  27.24 &    5.070 &   14.556 &    7.911 &   53.817 &   15.414 &   54.056 &   55.751 &   55.981 &   56.537 \\
SDSS J0926               &   2.08 &    0.686 &    1.185 &    0.649 &    4.852 &    1.369 &    4.900 &    4.994 &    5.041 &    5.083 \\
CP Eri                   &   3.66 &    0.871 &    2.842 &    1.560 &   11.291 &    2.973 &   11.325 &   11.643 &   11.676 &   11.780 \\
V406 Hya                 &   2.72 &    0.968 &    1.488 &    0.868 &    6.094 &    1.775 &    6.171 &    6.273 &    6.348 &    6.407 \\
\multicolumn{11}{c}{\textbf{\ac{DWD}}}\\
ZTF J1539                &  14.27 &   39.760 &  116.419 &  211.738 &  573.196 &  123.023 &  574.573 &  584.899 &  586.249 &  623.315 \\
ZTF J0546                &   7.78 &   42.847 &  100.572 &  162.448 &  315.746 &  109.319 &  318.640 &  331.376 &  334.135 &  371.531 \\
ZTF J1858                &   9.40 &   52.832 &  117.498 &  168.997 &  325.419 &  128.829 &  329.680 &  345.982 &  349.992 &  388.657 \\
ZTF J2243                &  10.27 &   20.003 &   78.786 &  110.980 &  231.563 &   81.286 &  232.426 &  244.599 &  245.416 &  269.343 \\
SDSS J0651               &  16.77 &   27.459 &   95.302 &   59.932 &  200.927 &   99.180 &  202.795 &  222.383 &  224.072 &  231.949 \\
ZTF J0538                &  16.39 &   18.951 &   75.748 &   39.118 &  155.205 &   78.083 &  156.358 &  172.704 &  173.740 &  178.089 \\
ZTF J1905                &  24.80 &   26.375 &   79.696 &   36.783 &  215.856 &   83.946 &  217.461 &  230.098 &  231.605 &  234.508 \\
SDSS J0935               &  48.71 &   46.878 &  102.361 &   48.164 &  308.927 &  112.585 &  312.463 &  325.443 &  328.802 &  332.311 \\
SDSS J2322               &  10.34 &   11.865 &   35.216 &   16.639 &   99.057 &   37.161 &   99.765 &  105.130 &  105.798 &  107.098 \\
PTF J0533                &   8.15 &    5.317 &   12.285 &    5.857 &   39.871 &   13.386 &   40.224 &   41.720 &   42.058 &   42.464 \\
ZTF J2029                &   8.37 &    4.761 &    9.200 &    4.408 &   30.947 &   10.358 &   31.311 &   32.285 &   32.634 &   32.931 \\
J1239                    &  15.89 &    5.351 &   19.498 &    9.664 &   68.357 &   20.219 &   68.566 &   71.083 &   71.284 &   71.937 \\
ZTF J0722                &   8.30 &    3.476 &    6.733 &    3.412 &   22.394 &    7.577 &   22.662 &   23.384 &   23.641 &   23.886 \\
ZTF J1749                &   6.85 &    2.172 &    4.673 &    2.489 &   15.897 &    5.153 &   16.045 &   16.570 &   16.712 &   16.896 \\
SDSS J0634               &  17.32 &   12.809 &   22.323 &   11.890 &   80.898 &   25.737 &   81.906 &   83.921 &   84.893 &   85.722 \\
ZTF J2228                &   5.93 &    1.529 &    4.448 &    2.442 &   18.736 &    4.704 &   18.798 &   19.257 &   19.317 &   19.471 \\
SMSS J0338               &  12.12 &    2.201 &    6.237 &    3.517 &   30.184 &    6.614 &   30.264 &   30.821 &   30.900 &   31.099 \\
ZTF J0643                &   5.07 &    1.505 &    2.378 &    1.433 &    9.407 &    2.814 &    9.527 &    9.703 &    9.819 &    9.923 \\
ZTF J0640                &   4.95 &    1.320 &    2.234 &    1.352 &    8.526 &    2.595 &    8.627 &    8.814 &    8.912 &    9.014 \\
SDSS J0106               &   7.61 &    0.908 &    2.889 &    1.775 &   11.410 &    3.028 &   11.446 &   11.770 &   11.805 &   11.938 \\
ZTF J2130                &   7.82 &    0.857 &    2.093 &    1.286 &    8.634 &    2.261 &    8.676 &    8.884 &    8.925 &    9.017 \\
SDSS J1630               &  13.89 &    1.709 &    5.168 &    3.188 &   23.976 &    5.443 &   24.037 &   24.527 &   24.586 &   24.792 \\
SDSS J0822               &  10.65 &    1.761 &    2.564 &    1.588 &   12.638 &    3.110 &   12.760 &   12.895 &   13.015 &   13.112 \\
J1526                    &  13.58 &    1.954 &    5.124 &    3.174 &   21.871 &    5.484 &   21.959 &   22.464 &   22.549 &   22.771 \\
ZTF J1901                &   6.46 &    0.607 &    1.530 &    0.951 &    7.184 &    1.646 &    7.210 &    7.345 &    7.370 &    7.431 \\
SDSS J1235               &  11.32 &    1.632 &    5.093 &    3.395 &   17.635 &    5.348 &   17.711 &   18.356 &   18.429 &   18.739 \\
SDSS J0056               &   9.17 &    0.470 &    1.830 &    1.322 &    6.001 &    1.889 &    6.019 &    6.273 &    6.291 &    6.428 \\
SDSS J1056               &   8.01 &    0.525 &    1.406 &    1.016 &    5.161 &    1.501 &    5.188 &    5.349 &    5.375 &    5.470 \\
SDSS J0923               &  27.25 &    2.513 &    4.576 &    3.350 &   16.417 &    5.220 &   16.608 &   17.043 &   17.227 &   17.550 \\
WD 0957                  &  25.51 &    0.498 &    2.077 &    1.680 &    7.881 &    2.136 &    7.897 &    8.150 &    8.165 &    8.336 \\
SDSS J1337               &  44.16 &    1.184 &    5.172 &    4.355 &   16.955 &    5.306 &   16.996 &   17.726 &   17.766 &   18.292 \\
WD 1242                  & 109.24 &    0.754 &    3.064 &    3.378 &   11.350 &    3.155 &   11.375 &   11.756 &   11.780 &   12.255 \\
WD 1704                  &  99.91 &    0.376 &    1.199 &    1.408 &    6.289 &    1.257 &    6.301 &    6.403 &    6.414 &    6.567 \\
\multicolumn{11}{c}{\textbf{\ac{CWDB}}}\\
ZTF J0526                &  18.16 &   12.780 &   35.825 &   17.056 &  106.417 &   38.036 &  107.182 &  112.285 &  113.010 &  114.290 \\
ZTF J1840                &   8.66 &    3.298 &    5.353 &    3.106 &   20.551 &    6.287 &   20.814 &   21.237 &   21.491 &   21.715 \\
ZTF J1707                &   4.94 &    1.476 &    2.933 &    1.709 &   11.425 &    3.283 &   11.520 &   11.796 &   11.888 &   12.010 \\
ZTF J2351                &   7.31 &    0.679 &    2.405 &    1.525 &   10.307 &    2.499 &   10.329 &   10.584 &   10.606 &   10.715 \\
ZTF J1611                &  33.09 &    2.285 &    8.172 &    5.444 &   36.843 &    8.485 &   36.914 &   37.738 &   37.808 &   38.197 \\
ZTF J1813                &  10.00 &    0.902 &    2.419 &    1.628 &   10.426 &    2.582 &   10.465 &   10.703 &   10.741 &   10.864 \\
ZTF J1603                &  20.10 &    0.665 &    2.431 &    1.892 &    8.045 &    2.520 &    8.072 &    8.404 &    8.430 &    8.640 \\
ZTF J0451                &  20.96 &    0.923 &    2.566 &    2.014 &    8.201 &    2.727 &    8.253 &    8.593 &    8.642 &    8.874 \\
ZTF J0112                &  16.83 &    0.433 &    1.909 &    1.501 &    6.374 &    1.957 &    6.389 &    6.654 &    6.668 &    6.835 \\
ZTF J2334                &  80.70 &    2.371 &    9.657 &    7.601 &   30.187 &    9.944 &   30.279 &   31.694 &   31.782 &   32.678 \\
ZTF J2007                & 135.63 &    7.199 &   15.212 &   12.003 &   50.028 &   16.829 &   50.543 &   52.289 &   52.782 &   54.130 \\
ZTF J1943                &  16.06 &    0.815 &    1.617 &    1.316 &    5.237 &    1.811 &    5.300 &    5.481 &    5.541 &    5.695 \\
ZTF J1708                &  52.55 &    2.044 &    5.611 &    4.576 &   16.984 &    5.972 &   17.106 &   17.886 &   18.003 &   18.575 \\
\multicolumn{11}{c}{\textbf{\ac{sdB}}}\\
OW J0741                 &   5.20 &    0.858 &    1.674 &    1.078 &    6.789 &    1.881 &    6.843 &    6.992 &    7.045 &    7.127 \\
CD-30                    &  33.07 &    0.909 &    3.340 &    2.513 &   12.491 &    3.461 &   12.524 &   12.930 &   12.962 &   13.203 \\
HD 265435                &  33.93 &    1.228 &    2.804 &    2.361 &    8.627 &    3.061 &    8.714 &    9.071 &    9.154 &    9.454 \\
\multicolumn{11}{c}{\textbf{\ac{UCXB}}}\\
1RXS J1718               &   4.85 &   31.404 &   65.188 &  118.093 &  256.238 &   72.358 &  258.155 &  264.400 &  266.258 &  291.272 \\
4U 1820-30               &   1.29 &    3.972 &   12.403 &    9.901 &   25.632 &   13.024 &   25.938 &   28.475 &   28.751 &   30.408 \\
4U 0513-40               &   4.56 &    3.614 &   15.016 &    6.949 &   43.255 &   15.445 &   43.406 &   45.788 &   45.930 &   46.453 \\
\end{longtable}

\end{widetext}

\vspace{1cm}

\section{parameters estimation of verification binaries}\label{app:estimate}

The estimation of parameters for the \acp{VB} is listed in 
Table~\ref{tb:parameter_estimation1}. 
The four key parameters are chirp mass, distance, inclination, and sky localization. 
Blank cells follow the rules explained in the main text and table caption: absence of a value means either the source was not detected by that instrument, or the inclination uncertainty exceeded 90$^\circ$.

\begin{widetext}
\setlength{\LTcapwidth}{\textwidth}
\begin{longtable}{ccccccccccc}
\caption{Uncertainties on $M_c$, $d$, $\iota$ and $\Omega$ for the 40 \acp{VB} with \ac{SNR} $\geq 20$ in the most complete network configuration, evaluated for each detector and for the full network. 
Blank entries indicate either that the source is not shown because its \ac{SNR} is below 20 in that configuration, or, for $\iota$ only, that the uncertainty exceeds 90$^\circ$ and is physically meaningless.}
\label{tb:parameter_estimation1}\\
\hline
\hline
\multicolumn{2}{c}{} & \multicolumn{9}{c}{$\Delta \theta$} \\ 
\cline{3-11} 
\multicolumn{1}{c}{Source} & \multicolumn{1}{c}{parameters} &
\multicolumn{1}{c}{TQ} & \multicolumn{1}{c}{LISA} & 
\multicolumn{1}{c}{Taiji} & \multicolumn{1}{c}{DECIGO} & \multicolumn{1}{c}{TL}  & 
\multicolumn{1}{c}{TD} & \multicolumn{1}{c}{LD} & \multicolumn{1}{c}{TLD} & \multicolumn{1}{c}{TLTD}\\
\hline
\hline
\endfirsthead
\multicolumn{11}{c}%
{\tablename\ \thetable\ -- \textit{Continued from previous page}} \\
\hline
\hline
\multicolumn{2}{c}{} & \multicolumn{9}{c}{$\Delta \theta$} \\ 
\cline{3-11} 
\multicolumn{1}{c}{Source} & \multicolumn{1}{c}{parameters} &
\multicolumn{1}{c}{TQ} & \multicolumn{1}{c}{LISA} & 
\multicolumn{1}{c}{Taiji} & \multicolumn{1}{c}{DECIGO} & \multicolumn{1}{c}{TL}  & 
\multicolumn{1}{c}{TD} & \multicolumn{1}{c}{LD} & \multicolumn{1}{c}{TLD} & \multicolumn{1}{c}{TLTD}\\
\hline
\hline
\endhead
\hline \multicolumn{11}{r}{\textit{Continued on next page}} \\
\endfoot
\hline
\endlastfoot
\multirow{4}{*}{J0806} & $M_c=3.31e-1$ ($\text{M}_\odot$) & 1.63e-3 & 1.10e-3 & 5.48e-4 & 1.65e-4 & 8.26e-4 & 4.22e-5 & 4.21e-5 & 4.18e-5 & 4.05e-5 \\
& $d=5.00e0$ (kpc) & 3.29e-1 & 3.49e-1 & 1.73e-1 & 3.95e-1 & 2.34e-1 & 1.79e-1 & 1.68e-1 & 1.37e-1 & 8.92e-2 \\
& $\iota=3.80e1$ (deg) & 5.40e0 & 5.66e0 & 2.82e0 & 7.74e0 & 3.82e0 & 3.31e0 & 3.10e0 & 2.49e0 & 1.58e0 \\
& $\Omega$ (deg$^2$) & 8.88e-3 & 4.37e-3 & 1.08e-3 & 2.49e-4 & 2.50e-3 & 1.11e-4 & 1.03e-4 & 8.70e-5 & 6.58e-5 \\
\hline
\multirow{4}{*}{ATLAS J1013} & $M_c=2.32e-1$ ($\text{M}_\odot$) & -- & 1.67e-2 & 1.14e-2 & 5.86e-3 & 1.61e-2 & 1.39e-3 & 1.34e-3 & 1.33e-3 & 1.23e-3 \\
& $d=1.80e0$ (kpc) & -- & 2.19e-1 & 1.49e-1 & 7.68e-2 & 2.10e-1 & 2.23e-2 & 2.13e-2 & 2.12e-2 & 1.94e-2 \\
& $\iota=8.20e1$ (deg) & -- & 7.21e-1 & 4.91e-1 & 2.14e-1 & 6.96e-1 & 2.13e-1 & 2.05e-1 & 2.04e-1 & 1.88e-1 \\
& $\Omega$ (deg$^2$) & -- & 1.34e-1 & 6.21e-2 & 1.46e-5 & 1.22e-2 & 1.46e-5 & 1.46e-5 & 1.45e-5 & 1.45e-5 \\
\hline
\multirow{4}{*}{V407 Vul} & $M_c=3.11e-1$ ($\text{M}_\odot$) & 4.37e-2 & 1.29e-2 & 1.05e-2 & 4.56e-3 & 1.22e-2 & 1.12e-3 & 1.07e-3 & 1.06e-3 & 9.98e-4 \\
& $d=2.09e0$ (kpc) & 5.08e-1 & 1.54e-1 & 1.25e-1 & 5.54e-2 & 1.45e-1 & 2.54e-2 & 2.37e-2 & 2.35e-2 & 2.15e-2 \\
& $\iota=6.00e1$ (deg) & 3.22e0 & 1.25e0 & 1.02e0 & 4.92e-1 & 1.15e0 & 4.86e-1 & 4.53e-1 & 4.48e-1 & 4.06e-1 \\
& $\Omega$ (deg$^2$) & 3.21e-4 & 9.95e-2 & 6.64e-2 & 3.07e-6 & 3.02e-4 & 3.03e-6 & 3.07e-6 & 3.03e-6 & 3.03e-6 \\
\hline
\multirow{4}{*}{ES Cet} & $M_c=2.95e-1$ ($\text{M}_\odot$) & -- & 1.49e-2 & 1.47e-2 & 6.01e-3 & 1.47e-2 & 1.65e-3 & 1.57e-3 & 1.56e-3 & 1.48e-3 \\
& $d=1.73e0$ (kpc) & -- & 1.56e-1 & 1.54e-1 & 6.42e-2 & 1.54e-1 & 2.78e-2 & 2.67e-2 & 2.58e-2 & 2.44e-2 \\
& $\iota=6.00e1$ (deg) & -- & 1.79e0 & 1.77e0 & 9.92e-1 & 1.74e0 & 7.27e-1 & 7.34e-1 & 6.50e-1 & 5.99e-1 \\
& $\Omega$ (deg$^2$) & -- & 2.00e-2 & 1.95e-2 & 1.13e-2 & 1.00e-2 & 6.60e-3 & 5.85e-3 & 4.35e-3 & 3.29e-3 \\
\hline
\multirow{4}{*}{ZTF J0127} & $M_c=3.15e-1$ ($\text{M}_\odot$) & -- & 9.69e-2 & 1.75e-1 & 4.34e-2 & 9.43e-2 & 9.96e-3 & 9.25e-3 & 9.23e-3 & 9.05e-3 \\
& $d=1.28e0$ (kpc) & -- & 6.59e-1 & 1.19e0 & 2.96e-1 & 6.42e-1 & 6.95e-2 & 6.46e-2 & 6.45e-2 & 6.32e-2 \\
& $\iota=8.00e1$ (deg) & -- & 7.48e-1 & 1.35e0 & 3.62e-1 & 7.42e-1 & 3.62e-1 & 3.24e-1 & 3.24e-1 & 3.14e-1 \\
& $\Omega$ (deg$^2$) & -- & 2.87e-1 & 9.35e-1 & 4.08e-5 & 1.69e-1 & 4.08e-5 & 4.08e-5 & 4.08e-5 & 4.08e-5 \\
\hline
\multirow{4}{*}{SDSS J1351} & $M_c=2.24e-1$ ($\text{M}_\odot$) & -- & 3.23e-1 & -- & 1.30e-1 & 3.20e-1 & 3.39e-2 & 3.21e-2 & 3.20e-2 & 3.17e-2 \\
& $d=1.34e0$ (kpc) & -- & 3.22e0 & -- & 1.37e0 & 3.19e0 & 4.77e-1 & 3.23e-1 & 3.23e-1 & 3.17e-1 \\
& $\iota=6.00e1$ (deg) & -- & 4.33e0 & -- & 4.18e1 & 4.28e0 & 2.35e1 & 3.56e0 & 3.54e0 & 3.24e0 \\
& $\Omega$ (deg$^2$) & -- & 3.70e-1 & -- & 9.16e-1 & 2.42e-1 & 3.00e-1 & 3.35e-2 & 3.19e-2 & 3.08e-2 \\
\hline
\multirow{4}{*}{AM CVn} & $M_c=2.38e-1$ ($\text{M}_\odot$) & 5.27e-1 & 8.14e-2 & 1.76e-1 & 3.24e-2 & 7.94e-2 & 8.66e-3 & 8.15e-3 & 8.13e-3 & 8.03e-3 \\
& $d=3.02e-1$ (kpc) & 1.12e0 & 1.73e-1 & 3.75e-1 & 6.85e-2 & 1.69e-1 & 1.92e-2 & 1.82e-2 & 1.81e-2 & 1.79e-2 \\
& $\iota=4.30e1$ (deg) & 1.38e1 & 3.49e0 & 7.55e0 & 1.43e0 & 3.35e0 & 1.42e0 & 1.31e0 & 1.30e0 & 1.28e0 \\
& $\Omega$ (deg$^2$) & 2.43e-4 & 4.58e-2 & 2.15e-1 & 4.71e-6 & 2.38e-4 & 4.59e-6 & 4.71e-6 & 4.59e-6 & 4.59e-6 \\
\hline
\multirow{4}{*}{SDSS J1908} & $M_c=2.04e-1$ ($\text{M}_\odot$) & -- & 6.11e-1 & -- & 1.93e-1 & 5.85e-1 & 4.60e-2 & 4.43e-2 & 4.42e-2 & 4.39e-2 \\
& $d=9.77e-1$ (kpc) & -- & 5.26e0 & -- & 1.73e0 & 5.04e0 & 8.88e-1 & 8.23e-1 & 8.17e-1 & 8.04e-1 \\
& $\iota=1.50e1$ (deg) & -- & -- & -- & -- & -- & -- & -- & -- & -- \\
& $\Omega$ (deg$^2$) & -- & 3.91e0 & -- & 1.82e-5 & 1.32e-2 & 1.81e-5 & 1.82e-5 & 1.81e-5 & 1.81e-5 \\
\hline
\multirow{4}{*}{HP Lib} & $M_c=1.64e-1$ ($\text{M}_\odot$) & -- & 3.02e-1 & 6.54e-1 & 1.07e-1 & 2.94e-1 & 2.76e-2 & 2.67e-2 & 2.62e-2 & 2.60e-2 \\
& $d=2.80e-1$ (kpc) & -- & 8.64e-1 & 1.87e0 & 3.45e-1 & 8.43e-1 & 1.15e-1 & 8.51e-2 & 8.22e-2 & 8.02e-2 \\
& $\iota=3.00e1$ (deg) & -- & 3.00e1 & 6.50e1 & 6.99e1 & 2.21e1 & 3.04e1 & 1.68e1 & 1.48e1 & 1.38e1 \\
& $\Omega$ (deg$^2$) & -- & 7.60e-2 & 3.56e-1 & 3.94e-2 & 1.56e-4 & 1.54e-4 & 5.55e-3 & 1.51e-4 & 1.51e-4 \\
\hline
\multirow{4}{*}{TIC378898110} & $M_c=2.24e-1$ ($\text{M}_\odot$) & -- & 1.65e0 & -- & 4.38e-1 & 1.57e0 & 9.91e-2 & 9.65e-2 & 9.63e-2 & 9.57e-2 \\
& $d=3.09e-1$ (kpc) & -- & 3.79e0 & -- & 1.01e0 & 3.61e0 & 2.28e-1 & 2.22e-1 & 2.21e-1 & 2.20e-1 \\
& $\iota=7.40e1$ (deg) & -- & 2.39e0 & -- & 6.30e-1 & 2.37e0 & 6.29e-1 & 6.08e-1 & 6.08e-1 & 6.03e-1 \\
& $\Omega$ (deg$^2$) & -- & 3.85e0 & -- & 2.80e-5 & 2.70e0 & 2.80e-5 & 2.80e-5 & 2.80e-5 & 2.80e-5 \\
\hline
\multirow{4}{*}{CXOGBS J1751} & $M_c=1.73e-1$ ($\text{M}_\odot$) & -- & -- & -- & 1.96e0 & -- & 4.68e-1 & 4.59e-1 & 4.54e-1 & 4.51e-1 \\
& $d=9.39e-1$ (kpc) & -- & -- & -- & 1.77e1 & -- & 4.23e0 & 4.13e0 & 4.09e0 & 4.06e0 \\
& $\iota=6.00e1$ (deg) & -- & -- & -- & 6.00e1 & -- & 1.63e1 & 1.82e1 & 1.29e1 & 1.23e1 \\
& $\Omega$ (deg$^2$) & -- & -- & -- & 5.28e0 & -- & 2.80e-3 & 9.50e-1 & 2.79e-3 & 2.79e-3 \\
\hline
\multirow{4}{*}{CR Boo} & $M_c=1.84e-1$ ($\text{M}_\odot$) & -- & 2.16e0 & -- & 5.93e-1 & 2.12e0 & 1.64e-1 & 1.59e-1 & 1.57e-1 & 1.56e-1 \\
& $d=3.52e-1$ (kpc) & -- & 6.92e0 & -- & 1.90e0 & 6.77e0 & 5.41e-1 & 5.17e-1 & 5.16e-1 & 5.10e-1 \\
& $\iota=3.00e1$ (deg) & -- & 8.93e1 & -- & 3.48e1 & 8.19e1 & 2.46e1 & 2.65e1 & 2.28e1 & 2.25e1 \\
& $\Omega$ (deg$^2$) & -- & 2.29e0 & -- & 5.10e-1 & 4.41e-2 & 3.93e-2 & 2.67e-1 & 3.63e-2 & 3.57e-2 \\
\hline
\multirow{4}{*}{V803 Cen} & $M_c=2.20e-1$ ($\text{M}_\odot$) & -- & 2.33e0 & -- & 6.02e-1 & 2.24e0 & 1.34e-1 & 1.30e-1 & 1.30e-1 & 1.29e-1 \\
& $d=2.87e-1$ (kpc) & -- & 5.07e0 & -- & 1.32e0 & 4.89e0 & 3.35e-1 & 3.28e-1 & 3.27e-1 & 3.25e-1 \\
& $\iota=1.35e1$ (deg) & -- & -- & -- & -- & -- & -- & -- & -- & -- \\
& $\Omega$ (deg$^2$) & -- & 2.64e-2 & -- & 7.68e-6 & 2.60e-2 & 7.68e-6 & 7.67e-6 & 7.67e-6 & 7.67e-6 \\
\hline
\multirow{4}{*}{SMSS J1138} & $M_c=4.05e-1$ ($\text{M}_\odot$) & -- & -- & -- & 9.12e-1 & -- & 2.18e-1 & 2.13e-1 & 2.13e-1 & 2.11e-1 \\
& $d=5.47e-1$ (kpc) & -- & -- & -- & 2.05e0 & -- & 4.92e-1 & 4.80e-1 & 4.79e-1 & 4.75e-1 \\
& $\iota=8.87e1$ (deg) & -- & -- & -- & 5.67e-1 & -- & 5.67e-1 & 5.51e-1 & 5.51e-1 & 5.46e-1 \\
& $\Omega$ (deg$^2$) & -- & -- & -- & 1.40e-2 & -- & 1.40e-2 & 1.39e-2 & 1.39e-2 & 1.38e-2 \\
\hline
\multirow{4}{*}{ZTF J1539} & $M_c=3.03e-1$ ($\text{M}_\odot$) & 1.27e-2 & 2.80e-3 & 1.54e-3 & 5.52e-4 & 2.70e-3 & 1.53e-4 & 1.51e-4 & 1.51e-4 & 1.45e-4 \\
& $d=1.63e0$ (kpc) & 1.34e-1 & 2.91e-2 & 1.60e-2 & 5.83e-3 & 2.79e-2 & 3.38e-3 & 3.31e-3 & 3.30e-3 & 3.08e-3 \\
& $\iota=8.40e1$ (deg) & 1.01e1 & 3.26e-1 & 1.79e-1 & 7.67e-2 & 3.25e-1 & 7.67e-2 & 7.47e-2 & 7.47e-2 & 6.89e-2 \\
& $\Omega$ (deg$^2$) & 1.06e-1 & 4.14e-2 & 1.25e-2 & 5.09e-6 & 2.65e-2 & 5.09e-6 & 5.08e-6 & 5.08e-6 & 5.06e-6 \\
\hline
\multirow{4}{*}{ZTF J0546} & $M_c=3.65e-1$ ($\text{M}_\odot$) & 1.70e-2 & 5.19e-3 & 3.21e-3 & 1.73e-3 & 4.81e-3 & 3.94e-4 & 3.88e-4 & 3.79e-4 & 3.50e-4 \\
& $d=3.71e0$ (kpc) & 3.53e-1 & 1.29e-1 & 8.00e-2 & 6.28e-2 & 1.15e-1 & 3.46e-2 & 3.72e-2 & 3.18e-2 & 2.70e-2 \\
& $\iota=6.00e1$ (deg) & 2.66e0 & 1.33e0 & 8.22e-1 & 1.21e0 & 1.10e0 & 5.50e-1 & 6.65e-1 & 4.98e-1 & 4.11e-1 \\
& $\Omega$ (deg$^2$) & 3.29e-5 & 3.80e-2 & 1.46e-2 & 8.57e-3 & 3.28e-5 & 3.24e-5 & 3.62e-3 & 3.23e-5 & 3.21e-5 \\
\hline
\multirow{4}{*}{ZTF J1858} & $M_c=3.65e-1$ ($\text{M}_\odot$) & 1.90e-2 & 6.10e-3 & 4.24e-3 & 2.14e-3 & 5.60e-3 & 5.34e-4 & 5.21e-4 & 5.12e-4 & 4.77e-4 \\
& $d=2.90e0$ (kpc) & 2.83e-1 & 1.07e-1 & 7.41e-2 & 4.08e-1 & 9.06e-2 & 7.53e-2 & 4.77e-2 & 4.19e-2 & 2.75e-2 \\
& $\iota=6.00e1$ (deg) & 2.14e0 & 1.39e0 & 9.64e-1 & 1.52e1 & 1.02e0 & 1.60e0 & 1.08e0 & 9.17e-1 & 5.99e-1 \\
& $\Omega$ (deg$^2$) & 4.14e-6 & 7.28e-3 & 3.52e-3 & 1.02e-2 & 4.14e-6 & 4.09e-6 & 3.57e-4 & 4.08e-6 & 4.07e-6 \\
\hline
\multirow{4}{*}{ZTF J2243} & $M_c=2.86e-1$ ($\text{M}_\odot$) & 6.18e-2 & 1.18e-2 & 8.39e-3 & 4.01e-3 & 1.14e-2 & 9.38e-4 & 9.01e-4 & 8.99e-4 & 8.38e-4 \\
& $d=1.75e0$ (kpc) & 6.38e-1 & 1.23e-1 & 8.73e-2 & 4.18e-2 & 1.19e-1 & 1.28e-2 & 1.22e-2 & 1.22e-2 & 1.12e-2 \\
& $\iota=8.21e1$ (deg) & 3.43e0 & 5.13e-1 & 3.64e-1 & 1.55e-1 & 5.05e-1 & 1.55e-1 & 1.49e-1 & 1.48e-1 & 1.37e-1 \\
& $\Omega$ (deg$^2$) & 2.25e-2 & 1.08e-1 & 5.42e-2 & 9.33e-6 & 1.38e-2 & 9.33e-6 & 9.32e-6 & 9.32e-6 & 9.31e-6 \\
\hline
\multirow{4}{*}{SDSS J0651} & $M_c=3.11e-1$ ($\text{M}_\odot$) & 1.67e-1 & 3.36e-2 & 5.34e-2 & 1.66e-2 & 3.23e-2 & 3.93e-3 & 3.73e-3 & 3.68e-3 & 3.60e-3 \\
& $d=9.58e-1$ (kpc) & 8.61e-1 & 1.73e-1 & 2.75e-1 & 8.72e-2 & 1.66e-1 & 2.26e-2 & 1.98e-2 & 1.96e-2 & 1.90e-2 \\
& $\iota=8.69e1$ (deg) & 1.07e0 & 5.97e-1 & 9.50e-1 & 4.71e0 & 4.49e-1 & 6.44e-1 & 4.16e-1 & 3.71e-1 & 3.31e-1 \\
& $\Omega$ (deg$^2$) & 4.79e-4 & 4.16e-2 & 1.05e-1 & 1.15e-1 & 4.65e-4 & 4.46e-4 & 6.54e-3 & 4.34e-4 & 4.32e-4 \\
\hline
\multirow{4}{*}{ZTF J0538} & $M_c=3.29e-1$ ($\text{M}_\odot$) & -- & 6.45e-2 & 1.25e-1 & 3.04e-2 & 6.29e-2 & 7.71e-3 & 7.26e-3 & 7.20e-3 & 7.09e-3 \\
& $d=9.97e-1$ (kpc) & -- & 3.26e-1 & 6.31e-1 & 1.62e-1 & 3.17e-1 & 4.34e-2 & 3.72e-2 & 3.70e-2 & 3.63e-2 \\
& $\iota=8.54e1$ (deg) & -- & 9.17e-1 & 1.78e0 & 9.55e0 & 7.49e-1 & 1.48e0 & 7.66e-1 & 6.78e-1 & 6.20e-1 \\
& $\Omega$ (deg$^2$) & -- & 6.06e-2 & 2.27e-1 & 4.86e-2 & 5.03e-3 & 2.89e-3 & 7.64e-3 & 2.30e-3 & 2.25e-3 \\
\hline
\multirow{4}{*}{ZTF J1905} & $M_c=3.65e-1$ ($\text{M}_\odot$) & 4.65e-1 & 1.24e-1 & 2.68e-1 & 4.40e-2 & 1.18e-1 & 1.05e-2 & 1.01e-2 & 1.00e-2 & 9.95e-3 \\
& $d=6.96e-1$ (kpc) & 1.48e0 & 3.94e-1 & 8.54e-1 & 1.40e-1 & 3.77e-1 & 3.46e-2 & 3.31e-2 & 3.30e-2 & 3.27e-2 \\
& $\iota=6.00e1$ (deg) & 4.38e0 & 1.52e0 & 3.29e0 & 5.44e-1 & 1.43e0 & 5.39e-1 & 5.11e-1 & 5.07e-1 & 5.01e-1 \\
& $\Omega$ (deg$^2$) & 1.11e-3 & 4.41e-1 & 2.07e0 & 1.92e-6 & 1.06e-3 & 1.92e-6 & 1.92e-6 & 1.92e-6 & 1.92e-6 \\
\hline
\multirow{4}{*}{SDSS J0935} & $M_c=4.13e-1$ ($\text{M}_\odot$) & 4.08e-1 & 1.25e-1 & 2.65e-1 & 4.47e-2 & 1.13e-1 & 9.96e-3 & 9.62e-3 & 9.54e-3 & 9.46e-3 \\
& $d=3.96e-1$ (kpc) & 6.51e-1 & 1.99e-1 & 4.23e-1 & 7.13e-2 & 1.81e-1 & 1.60e-2 & 1.55e-2 & 1.54e-2 & 1.52e-2 \\
& $\iota=6.00e1$ (deg) & 2.42e0 & 1.29e0 & 2.75e0 & 3.92e-1 & 1.08e0 & 3.87e-1 & 3.73e-1 & 3.68e-1 & 3.65e-1 \\
& $\Omega$ (deg$^2$) & 4.85e-5 & 1.61e-2 & 7.27e-2 & 1.11e-6 & 4.74e-5 & 1.07e-6 & 1.11e-6 & 1.07e-6 & 1.07e-6 \\
\hline
\multirow{4}{*}{SDSS J2322} & $M_c=2.61e-1$ ($\text{M}_\odot$) & -- & 5.41e-1 & -- & 1.98e-1 & 5.15e-1 & 4.99e-2 & 4.80e-2 & 4.77e-2 & 4.73e-2 \\
& $d=8.60e-1$ (kpc) & -- & 2.98e0 & -- & 1.30e0 & 2.85e0 & 5.57e-1 & 3.89e-1 & 3.69e-1 & 3.55e-1 \\
& $\iota=2.70e1$ (deg) & -- & 7.10e1 & -- & -- & 6.06e1 & 7.14e1 & 4.41e1 & 3.99e1 & 3.77e1 \\
& $\Omega$ (deg$^2$) & -- & 5.73e-1 & -- & 2.25e-1 & 1.53e-2 & 1.30e-2 & 4.58e-2 & 6.65e-3 & 6.32e-3 \\
\hline
\multirow{4}{*}{PTF J0533} & $M_c=2.75e-1$ ($\text{M}_\odot$) & -- & -- & -- & 4.71e-1 & -- & 1.28e-1 & 1.27e-1 & 1.24e-1 & 1.23e-1 \\
& $d=1.17e0$ (kpc) & -- & -- & -- & 3.34e0 & -- & 9.09e-1 & 9.03e-1 & 8.78e-1 & 8.72e-1 \\
& $\iota=7.28e1$ (deg) & -- & -- & -- & 2.92e0 & -- & 1.70e0 & 1.98e0 & 1.56e0 & 1.54e0 \\
& $\Omega$ (deg$^2$) & -- & -- & -- & 1.18e0 & -- & 4.80e-3 & 6.07e-1 & 4.76e-3 & 4.76e-3 \\
\hline
\multirow{4}{*}{ZTF J2029} & $M_c=2.70e-1$ ($\text{M}_\odot$) & -- & -- & -- & 6.85e-1 & -- & 1.78e-1 & 1.75e-1 & 1.73e-1 & 1.72e-1 \\
& $d=1.09e0$ (kpc) & -- & -- & -- & 4.63e0 & -- & 1.21e0 & 1.18e0 & 1.17e0 & 1.17e0 \\
& $\iota=8.66e1$ (deg) & -- & -- & -- & 1.06e0 & -- & 1.05e0 & 1.01e0 & 9.94e-1 & 9.82e-1 \\
& $\Omega$ (deg$^2$) & -- & -- & -- & 4.50e-3 & -- & 4.34e-3 & 4.48e-3 & 4.32e-3 & 4.31e-3 \\
\hline
\multirow{4}{*}{J1239} & $M_c=3.81e-1$ ($\text{M}_\odot$) & -- & -- & -- & 3.58e-1 & -- & 8.79e-2 & 8.53e-2 & 8.50e-2 & 8.43e-2 \\
& $d=9.72e-1$ (kpc) & -- & -- & -- & 1.52e0 & -- & 3.75e-1 & 3.65e-1 & 3.62e-1 & 3.60e-1 \\
& $\iota=7.10e1$ (deg) & -- & -- & -- & 3.51e0 & -- & 3.02e0 & 2.10e0 & 1.97e0 & 1.87e0 \\
& $\Omega$ (deg$^2$) & -- & -- & -- & 5.03e-1 & -- & 2.21e-1 & 2.33e-1 & 1.26e-1 & 1.17e-1 \\
\hline
\multirow{4}{*}{ZTF J0722} & $M_c=3.08e-1$ ($\text{M}_\odot$) & -- & -- & -- & 1.38e0 & -- & 3.50e-1 & 3.42e-1 & 3.39e-1 & 3.37e-1 \\
& $d=1.27e0$ (kpc) & -- & -- & -- & 9.47e0 & -- & 2.40e0 & 2.34e0 & 2.33e0 & 2.31e0 \\
& $\iota=8.97e1$ (deg) & -- & -- & -- & 1.39e0 & -- & 1.37e0 & 1.33e0 & 1.31e0 & 1.30e0 \\
& $\Omega$ (deg$^2$) & -- & -- & -- & 6.08e-1 & -- & 5.82e-1 & 5.79e-1 & 5.54e-1 & 5.48e-1 \\
\hline
\multirow{4}{*}{ZTF J1749} & $M_c=2.90e-1$ ($\text{M}_\odot$) & -- & -- & -- & -- & -- & -- & 7.62e-1 & 7.57e-1 & 7.51e-1 \\
& $d=1.29e0$ (kpc) & -- & -- & -- & -- & -- & -- & 5.64e0 & 5.61e0 & 5.57e0 \\
& $\iota=8.55e1$ (deg) & -- & -- & -- & -- & -- & -- & 2.06e0 & 2.04e0 & 2.03e0 \\
& $\Omega$ (deg$^2$) & -- & -- & -- & -- & -- & -- & 1.00e-2 & 9.75e-3 & 9.74e-3 \\
\hline
\multirow{4}{*}{SDSS J0634} & $M_c=2.64e-1$ ($\text{M}_\odot$) & -- & 2.35e0 & -- & 6.68e-1 & 2.06e0 & 1.53e-1 & 1.51e-1 & 1.49e-1 & 1.48e-1 \\
& $d=4.35e-1$ (kpc) & -- & 6.46e0 & -- & 1.84e0 & 5.67e0 & 4.21e-1 & 4.18e-1 & 4.08e-1 & 4.05e-1 \\
& $\iota=4.30e1$ (deg) & -- & 2.45e1 & -- & 1.41e1 & 1.64e1 & 7.21e0 & 8.80e0 & 6.69e0 & 6.56e0 \\
& $\Omega$ (deg$^2$) & -- & 2.45e0 & -- & 6.53e-1 & 1.91e-4 & 1.91e-4 & 2.52e-1 & 1.91e-4 & 1.91e-4 \\
\hline
\multirow{4}{*}{SMSS J0338} & $M_c=2.56e-1$ ($\text{M}_\odot$) & -- & -- & -- & 2.96e0 & -- & 7.07e-1 & 6.97e-1 & 6.96e-1 & 6.92e-1 \\
& $d=5.36e-1$ (kpc) & -- & -- & -- & 1.03e1 & -- & 2.47e0 & 2.43e0 & 2.43e0 & 2.42e0 \\
& $\iota=6.90e1$ (deg) & -- & -- & -- & 2.34e0 & -- & 2.33e0 & 2.29e0 & 2.29e0 & 2.27e0 \\
& $\Omega$ (deg$^2$) & -- & -- & -- & 2.65e-4 & -- & 2.65e-4 & 2.65e-4 & 2.65e-4 & 2.65e-4 \\
\hline
\multirow{4}{*}{SDSS J1630} & $M_c=4.06e-1$ ($\text{M}_\odot$) & -- & -- & -- & 7.35e0 & -- & 1.98e0 & 1.95e0 & 1.95e0 & 1.94e0 \\
& $d=8.48e-1$ (kpc) & -- & -- & -- & 2.56e1 & -- & 6.92e0 & 6.81e0 & 6.80e0 & 6.75e0 \\
& $\iota=6.00e1$ (deg) & -- & -- & -- & 6.01e0 & -- & 6.00e0 & 5.81e0 & 5.79e0 & 5.72e0 \\
& $\Omega$ (deg$^2$) & -- & -- & -- & 2.50e-4 & -- & 2.50e-4 & 2.50e-4 & 2.50e-4 & 2.50e-4 \\
\hline
\multirow{4}{*}{J1526} & $M_c=3.35e-1$ ($\text{M}_\odot$) & -- & -- & -- & 9.76e0 & -- & 2.37e0 & 2.32e0 & 2.31e0 & 2.29e0 \\
& $d=6.25e-1$ (kpc) & -- & -- & -- & 3.04e1 & -- & 7.35e0 & 7.20e0 & 7.16e0 & 7.10e0 \\
& $\iota=6.00e1$ (deg) & -- & -- & -- & 3.68e1 & -- & 1.42e1 & 1.24e1 & 1.10e1 & 1.03e1 \\
& $\Omega$ (deg$^2$) & -- & -- & -- & 9.57e0 & -- & 3.01e-2 & 1.96e0 & 2.97e-2 & 2.96e-2 \\
\hline
\multirow{4}{*}{ZTF J0526} & $M_c=3.65e-1$ ($\text{M}_\odot$) & -- & 4.88e-1 & -- & 1.75e-1 & 4.56e-1 & 3.65e-2 & 3.52e-2 & 3.50e-2 & 3.47e-2 \\
& $d=8.45e-1$ (kpc) & -- & 1.88e0 & -- & 6.74e-1 & 1.76e0 & 1.42e-1 & 1.37e-1 & 1.37e-1 & 1.35e-1 \\
& $\iota=6.00e1$ (deg) & -- & 3.57e0 & -- & 1.15e0 & 3.30e0 & 1.14e0 & 1.08e0 & 1.08e0 & 1.06e0 \\
& $\Omega$ (deg$^2$) & -- & 5.30e-1 & -- & 1.67e-5 & 1.03e-3 & 1.64e-5 & 1.67e-5 & 1.64e-5 & 1.64e-5 \\
\hline
\multirow{4}{*}{ZTF J1840} & $M_c=3.65e-1$ ($\text{M}_\odot$) & -- & -- & -- & 4.57e0 & -- & 1.16e0 & 1.14e0 & 1.13e0 & 1.12e0 \\
& $d=1.29e0$ (kpc) & -- & -- & -- & 2.69e1 & -- & 6.79e0 & 6.70e0 & 6.62e0 & 6.57e0 \\
& $\iota=6.00e1$ (deg) & -- & -- & -- & -- & -- & 1.81e1 & 1.94e1 & 1.47e1 & 1.39e1 \\
& $\Omega$ (deg$^2$) & -- & -- & -- & 2.11e1 & -- & 3.13e-3 & 1.23e0 & 3.13e-3 & 3.13e-3 \\
\hline
\multirow{4}{*}{ZTF J1611} & $M_c=3.65e-1$ ($\text{M}_\odot$) & -- & -- & -- & 1.05e1 & -- & 2.80e0 & 2.75e0 & 2.75e0 & 2.73e0 \\
& $d=2.57e-1$ (kpc) & -- & -- & -- & 1.24e1 & -- & 3.28e0 & 3.23e0 & 3.22e0 & 3.20e0 \\
& $\iota=6.00e1$ (deg) & -- & -- & -- & 3.21e0 & -- & 3.20e0 & 3.13e0 & 3.13e0 & 3.10e0 \\
& $\Omega$ (deg$^2$) & -- & -- & -- & 1.41e-4 & -- & 1.41e-4 & 1.41e-4 & 1.41e-4 & 1.41e-4 \\
\hline
\multirow{4}{*}{ZTF J2334} & $M_c=3.65e-1$ ($\text{M}_\odot$) & -- & -- & -- & 9.62e1 & -- & 2.21e1 & 2.13e1 & 2.13e1 & 2.09e1 \\
& $d=7.60e-2$ (kpc) & -- & -- & -- & 3.34e1 & -- & 7.68e0 & 7.41e0 & 7.39e0 & 7.24e0 \\
& $\iota=6.00e1$ (deg) & -- & -- & -- & 4.11e0 & -- & 4.10e0 & 3.90e0 & 3.89e0 & 3.77e0 \\
& $\Omega$ (deg$^2$) & -- & -- & -- & 2.05e-4 & -- & 2.05e-4 & 2.05e-4 & 2.05e-4 & 2.05e-4 \\
\hline
\multirow{4}{*}{ZTF J2007} & $M_c=3.65e-1$ ($\text{M}_\odot$) & -- & -- & -- & 5.58e1 & -- & 1.38e1 & 1.34e1 & 1.33e1 & 1.31e1 \\
& $d=4.50e-2$ (kpc) & -- & -- & -- & 1.15e1 & -- & 2.83e0 & 2.75e0 & 2.74e0 & 2.69e0 \\
& $\iota=6.00e1$ (deg) & -- & -- & -- & 2.45e0 & -- & 2.42e0 & 2.33e0 & 2.30e0 & 2.23e0 \\
& $\Omega$ (deg$^2$) & -- & -- & -- & 6.67e-5 & -- & 6.41e-5 & 6.67e-5 & 6.41e-5 & 6.41e-5 \\
\hline
\multirow{4}{*}{1RXS J1718} & $M_c=3.65e-1$ ($\text{M}_\odot$) & 1.42e-2 & 4.97e-3 & 2.74e-3 & 1.32e-3 & 4.52e-3 & 2.99e-4 & 2.97e-4 & 2.92e-4 & 2.73e-4 \\
& $d=6.50e0$ (kpc) & 6.46e-1 & 2.94e-1 & 1.62e-1 & 1.14e-1 & 2.56e-1 & 6.90e-2 & 8.40e-2 & 6.55e-2 & 5.72e-2 \\
& $\iota=6.00e1$ (deg) & 3.65e0 & 1.97e0 & 1.09e0 & 1.20e0 & 1.67e0 & 6.21e-1 & 8.63e-1 & 5.84e-1 & 5.00e-1 \\
& $\Omega$ (deg$^2$) & 1.38e-4 & 2.73e-2 & 8.33e-3 & 5.20e-3 & 1.38e-4 & 1.33e-4 & 3.45e-3 & 1.32e-4 & 1.28e-4 \\
\hline
\multirow{4}{*}{4U 1820-30} & $M_c=2.28e-1$ ($\text{M}_\odot$) & -- & -- & -- & 1.06e-1 & -- & 2.53e-2 & 2.38e-2 & 2.35e-2 & 2.26e-2 \\
& $d=7.97e0$ (kpc) & -- & -- & -- & 7.61e0 & -- & 2.11e0 & 1.77e0 & 1.64e0 & 1.50e0 \\
& $\iota=6.00e1$ (deg) & -- & -- & -- & 5.16e1 & -- & 1.23e1 & 1.10e1 & 8.33e0 & 7.20e0 \\
& $\Omega$ (deg$^2$) & -- & -- & -- & 1.44e0 & -- & 1.16e-3 & 2.68e-1 & 1.16e-3 & 1.16e-3 \\
\hline
\multirow{4}{*}{4U 0513-40} & $M_c=5.00e-1$ ($\text{M}_\odot$) & -- & -- & -- & 1.64e-1 & -- & 4.25e-2 & 4.08e-2 & 4.07e-2 & 4.04e-2 \\
& $d=6.47e0$ (kpc) & -- & -- & -- & 3.56e0 & -- & 1.01e0 & 9.61e-1 & 9.58e-1 & 9.48e-1 \\
& $\iota=6.00e1$ (deg) & -- & -- & -- & 3.38e0 & -- & 3.36e0 & 3.10e0 & 3.09e0 & 3.04e0 \\
& $\Omega$ (deg$^2$) & -- & -- & -- & 9.37e-5 & -- & 9.35e-5 & 9.36e-5 & 9.34e-5 & 9.34e-5 \\
\end{longtable}

\end{widetext}

\bibliography{document}
\end{document}